\documentclass[11pt]{article}

\usepackage{verbatim}
\usepackage{amsmath}
\usepackage{amsfonts}
\usepackage{amssymb}
\usepackage{bbm}
\usepackage{bbold}
\usepackage{latexsym}
\usepackage{rotating} 
\usepackage{hhline} 
\usepackage{epsfig}
\usepackage[usenames,dvipsnames]{color}
\usepackage{graphicx}
\usepackage{enumitem}
\usepackage[normalem]{ulem} 
\usepackage[table]{xcolor}
\usepackage{subfig}
\usepackage{multirow}
\usepackage{mathrsfs}
\usepackage{blkarray}
\usepackage{amscd}
\usepackage{cite}
\usepackage{arydshln}

\usepackage{tikz}
\usetikzlibrary{arrows.meta} 

\DeclareMathSymbol{\mg}{\mathrel}{symbols}{"1D}

\newcommand{\bes}{\begin{split}}
\newcommand{\ees}{\end{split}}
\renewcommand{\arraystretch}{1.5}

\newcommand{\ga}{\alpha}
\newcommand{\gb}{\beta}
\renewcommand{\gg}{\gamma}
\newcommand{\gd}{\delta}

\newcommand{\gf}{\phi}

\newcommand{\gm}{\mu}
\newcommand{\gn}{\nu}

\newcommand{\gr}{\rho}

\newcommand{\gp}{\pi}

\newcommand{\gD}{\Delta}

\newcommand{\cF}{{\cal F}}

\newcommand{\cN}{{\cal N}}

\newcommand{\cR}{{\cal R}}

\newcommand{\cV}{{\cal V}}

\newcommand{\tr}{\text{tr}}

\newcommand{\Id}{\text{\small 1}\hspace{-3.5pt}\text{1}}

\newcommand{\ra}{\rightarrow}

\newcommand{\dsp}{\displaystyle}

\newcommand{\undr}[1]{{\underline{#1}}}

\newcommand{\beq}{\begin{equation}}
\newcommand{\eeq}{\end{equation}}
\newcommand{\barr}{\begin{array}}
\newcommand{\earr}{\end{array}}
\newcommand{\equ}[1]{\begin{gather} #1 \end{gather}}
\newcommand{\equa}[1]{\begin{align} #1 \end{align}}

\newcommand{\arry}[2]{\begin{array}{#1} #2 \end{array}}

\newcommand{\non}{\nonumber}
\newcommand{\sfrac}[2]{\mbox{$\frac{#1}{#2}$}}
\newcounter{oldcounter}

\newcommand{\Intr}{\mathbbm{Z}}
\newcommand{\Cplx}{\mathbbm{C}}

\newcommand{\ba}[2]{\[\begin{array}{#2}\label{#1}}
\newcommand{\ea}{\end{array}\]}
\newcommand{\be}{\begin{equation}}
\newcommand{\ee}{\end{equation}}
\newcommand{\bea}{\begin{eqnarray}}
\newcommand{\eea}{\end{eqnarray}}

\newcommand{\rep}[1]{\mathbf{#1}}
\newcommand{\crep}[1]{\overline{\rep{#1}}}

\newcommand{\sm}{{\,\mbox{-}}}

\newcommand{\ztwo}{\Intr_2\!\times\!\Intr_2}

\begin{document}

\thispagestyle{empty}

\begin{flushright}
LTH-1269\\ 
\end{flushright}
\vskip 1 cm
\begin{center}
{\Large {\bf 
Taming Triangulation Dependence of $\boldsymbol{T^6/\Intr_2\times\Intr_2}$ Resolutions
} 
}
\\[0pt]

\bigskip
\bigskip {\large
{\bf A.E.~Faraggi$^{a,}$}\footnote{
E-mail: alon.faraggi@liverpool.ac.uk},
{\bf S.~Groot Nibbelink$^{b,}$}\footnote{
E-mail: s.groot.nibbelink@hr.nl},
{\bf   M.~Hurtado Heredia$^{a,}$}\footnote{
E-mail: martin.hurtado@liv.ac.uk}
\bigskip }\\[0pt]
\vspace{0.23cm}
${}^a$ {\it 
Department of Mathematical Sciences, University of Liverpool, Liverpool L69 7ZL, UK 
 \\[1ex] } 
${}^b$ {\it 
School of Engineering and Applied Sciences, Rotterdam University of Applied Sciences, \\ 
G.J.\ de Jonghweg 4 - 6, 3015 GG Rotterdam, the Netherlands
 \\[1ex]
Research Centre Innovations in Care, Rotterdam University of Applied Sciences, \\ 
Postbus 25035, 3001 HA Rotterdam, the Netherlands
\\[1ex] 
School of Education, Rotterdam University of Applied Sciences, \\ 
Museumpark 40, 3015 CX Rotterdam, the Netherlands 
 } 
\\[1ex] 
\bigskip
\end{center}

\subsection*{\centering Abstract}

Resolutions of certain toroidal orbifolds, like $T^6/\Intr_2\times \Intr_2$, are far from unique, due to triangulation dependence of their resolved local singularities. This leads to an explosion of the number of topologically distinct smooth geometries associated to a single orbifold. By introducing a parameterisation to keep track of the triangulations used at all resolved singularities simultaneously, (self--)intersection numbers and integrated Chern classes can be determined for any triangulation configuration. Using this method the consistency conditions of line bundle models and the resulting chiral spectra can be worked out for any choice of triangulation. Moreover, by superimposing the Bianchi identities for all triangulation options much simpler though stronger conditions are uncovered.  When these are satisfied, flop--transitions between all different triangulations are admissible. Various methods are exemplified by a number of concrete models on resolutions of the $T^6/\Intr_2\times\Intr_2$ orbifold.

\newpage 
\setcounter{page}{1}

\section{Introduction}
\label{sc:Introduction}

String theory provides a perturbatively consistent approach to quantum gravity. An important advantage of string theory is that its consistency requirements mandate the existence of the gauge and matter structures that form the bedrock of the Standard Model of particle physics. As such, it enables the construction of phenomenological models, which in turn can be used to explore the theory and its possible relevance to observational data. Its internal consistency predicts the existence of a specific number of extra quantum fields propagating on a two dimensional string worldsheet,  which in some guise can be interpreted as extra spacetime dimensions beyond those observed in the physical world. Therefore, it has been suggested that these extra dimensions are compactified and are made sufficiently small to evade detection in contemporary experiments.

Phenomenological string models can be constructed by using exact worldsheet formulations of string theory in four dimensions, as well as target space tools that describe the effective field theory limit of string compactifications. Ultimately, a viable string theory model should have a low energy effective field theory description. Conversely, an effective field theory representation, which is compatible with string quantum gravity, should have a consistent ultra--violet embedding in string theory. However, at present the relation between these different regimes is poorly understood. The study of the consistency constraints on effective field theories of quantum gravity is a subject of intense contemporary research in the so--called ``Swampland program'' (for review and references see {\it e.g.}~\cite{Palti:2019pca}).

An alternative route is to explore the effective field theory limit of exact string solutions. This is hampered by the poor understanding of the moduli spaces of generic string compactifications.
Exact string solutions are typically studied by constructing the one--loop partition function and requiring it to be invariant under modular transformations. A plausible way forward is therefore to seek the imprint of the modular properties of the partition function in the effective field theory limit and their phenomenological  consequences in string models. $\Intr_2\times \Intr_2$ orbifolds of a six dimensional torus $T^6$ within the compactified heterotic--string are probably the most frequently studied examples of this route. Such compactifications have been analysed by using the free fermionic formulation~\cite{Antoniadis1987a,AB,Kawai1987} and the free bosonic formulation~\cite{Dixon:1985jw, Dixon:1986jc} of the heterotic--string in four dimensions. These free bosonic and fermionic worldsheet constructions are merely different languages to study the same physical object; a detailed dictionary can be employed to translate the models between the two descriptions~\cite{Athanasopoulos:2016aws}. Both languages were used to construct models that mimic the structure of the Minimal Supersymmetric Standard Model, {\it e.g.}~\cite{Faraggi:1989ka,Faraggi:1991jr,Faraggi:1992fa,Cleaver1999,Faraggi:2006qa,Faraggi:2017cnh} provide examples of free fermionic models and~\cite{Lebedev:2006kn,Lebedev:2007hv,Lebedev:2008un,Blaszczyk:2009in} of free bosonic constructions. Even though orbifolds are singular spaces, many quantities, like the full partition function, can be computed exactly at the one--loop level (and partially beyond) because of the power of the underlying modular symmetries.

However, these free worldsheet descriptions typically only apply to very specific points in the string moduli space or, at best, only parameterise a very small portion of the entire moduli space. In particular, to exploit the richness of the moduli space beyond the target space singularities, these singularities need to be deformed and/or resolved to form smooth Calabi--Yau compactifications with vector bundles~\cite{Candelas:1985en}. A variety of effective field theory and cohomology methods have been developed to study the resulting theories~\cite{Donagi:1998xe,Donagi:2000zs,Braun:2005ux,Bouchard:2005ag,Honecker:2006qz,Blumenhagen:2005ga,Anderson:2007nc,Anderson:2008uw,Anderson:2009mh,Nibbelink:2015ixa,GrootNibbelink:2015lme}. In particular, methods to resolve orbifold singularities using well--established toric geometry methodology have been worked out in many cases~\cite{Reffert:2006du,Nibbelink:2007rd,Nibbelink:2007pn,Nibbelink:2007xq,Nibbelink:2008tv,Lust:2006zh,Nibbelink:2009sp,Blaszczyk:2010db}.

The analysis of the effective field theory limit of $\Intr_2\times \Intr_2$ heterotic--string orbifolds and their resolutions is therefore well motivated from the phenomenological as well as the mathematical point of views. The analysis proceeds by the construction of toroidal $T^6/\Intr_2\times \Intr_2$ heterotic--string orbifolds and resolving the orbifold singularities using these well--established methodologies. However, a problematic caveat is the enormous number of possibilities that this opens up~\cite{Lust:2006zh,Nibbelink:2009sp,Blaszczyk:2010db}: The $T^6/\Intr_2\times \Intr_2$ orbifold has 64 $\Cplx^3/\Intr_2 \times \Intr_2$ singularities where $\Intr_2$--fixed tori intersect, which all need to be resolved to obtain a smooth geometry. Each $\Cplx^3/\Intr_2 \times \Intr_2$ singularity can be blown up in four topologically distinct ways encoded by four triangulations of the toric diagram of the resolved singularity. This results in a total of $4^{64}$ a priori distinct possibilities. While the symmetry structure of the $\Intr_2\times \Intr_2$ orbifold can be used to reduce this number by some factor, it still leaves a huge number (of the order of $10^{33}$) genuinely distinct choices. This is not a minor complication, as many physical properties of the resulting effective field theories are sensitively dependent on the  triangulation chosen. These range from the spectra of massless states in the low energy effective theory to the structure and strength of interactions among them. The only way to overcome this complication was by side stepping it: one simply makes some choice for the triangulation of all these resolved singularities and analyses the resulting physics in that particular case. This led to some insights in the structure of the theory in a somewhat larger part of the moduli space, but it seemed hopeless to extract any meaningful generic information about the properties of resolved $T^6/\Intr_2\times\Intr_2$ orbifolds.

A way forward is therefore to develop a formalism which allows computations for any choice of the triangulation of the 64 resolved $\Intr_2\times \Intr_2$ singularities. This is the task that we undertake in this paper. Moreover, having established such a method opens up the possibility to study some properties of resolved $T^6/\Intr_2\times \Intr_2$ orbifolds which are independent of triangulation choices or that hold in all possible triangulations simultaneously. To this end the paper has been structured as follows:

\subsubsection*{Outline}

Section~\ref{sc:ResT6Z2Z2} lays the foundation of this work by first recalling some basic facts of resolutions of the $T^6/\Intr_2\times \Intr_2$ orbifold and line bundle backgrounds on them. After that notation is developed to parameterise the triangulation choice at each of the 64 resolved $\Intr_2\times\Intr_2$ singularities, in terms of which the fundamental (self--)intersection numbers and the Chern classes are expressed. This allows to obtain relatively compact expressions for the volumes of curves, divisors and the manifold as a whole. Moreover, the flux quantisation conditions, the Bianchi identities and the multiplicity operator to determine the chiral spectrum can all be written down for any triangulation choice.

In Section~\ref{sc:TriangulationIndependence} it is argued that the flux quantisation conditions are, in fact, triangulation independent: if satisfied in a particular choice of triangulation, it holds for all. In addition, having written down Bianchi identities for any possible choice of triangulation of all 64 resolved singularities, one may wonder what requirements are obtained if one insists that these conditions hold for all triangulation choices simultaneously. Surprisingly, it can be shown that the resulting conditions are much simpler than those in any particular triangulation.

The following two sections provide various examples of the general results of the preceding two. In Section~\ref{sc:NoWilsonLines} models are considered without any Wilson lines so that all 64 resolved singularities may be treated in the same way. In particular, it stresses that the flux quantisation conditions are essential: when violated, the difference between the local multiplicities is not integral. Finally, Section~\ref{sc:BlasczcykGUT} revisits the so--called resolved Blaszczyk GUT model~\cite{Blaszczyk:2009in,Blaszczyk:2010db}. A model inspired by this GUT model is considered, which is consistent for any possible choice of triangulation.

The paper is completed with a summary and an outlook. The Appendix~\ref{sc:MathRelations} provides some useful identities for second and third Chern classes for manifolds with vanishing first Chern class.

\section{Resolutions of $\boldsymbol{T^6/\Intr_2\times\Intr_2}$}
\label{sc:ResT6Z2Z2} 

This section is devoted to develop some of the topological and geometrical properties of  resolutions of the toroidal orbifold $T^6/\Intr_2\times \Intr_2$. In fact, there are various $T^6/\Intr_2\times\Intr_2$ orbifolds~\cite{Donagi2004,Donagi2008,FRTV,Athanasopoulos:2016aws}: here we focus exclusively on the orbifold with Hodge numbers (51,3). Techniques to determine resolutions of toroidal orbifolds have been well--studied~\cite{Lust:2006zh}; here, in particular, the methods exploited in~\cite{Blaszczyk:2010db} are used. Also the resolutions of this orbifold have been considered before, however in the past one always had to make some assumptions which triangulation(s) to be considered, as the total number of choices (naively $4^{64}$) is a daunting number.  This section provides a brief review of this literature, but the main purpose is to develop a formalism to treat all of these possible triangulations simultaneously.

\subsection{The $\boldsymbol{T^6/\Intr_2\times \Intr_2}$ orbifold} 
\label{sc:T6Z22orbifold} 

The orbifold geometry will be taken to be factorisable of $T^6$ on the simplest rectangular lattice. The six torus coordinates are grouped into three complex ones on which two order--two orbifold reflections $R_1$, $R_2$ and their product $R_3=R_1R_2$ act. They are representations of $\Intr_2\times \Intr_2$ with non--trivial elements
\equ{
\text{diag}(R_1) = (1, -1, -1)~, 
\quad
\text{diag}(R_2) = (-1, 1, -1)~, 
\quad 
\text{diag}(R_3) = \text{diag}(R_1 R_2) = (-1, -1, 1)~. 
}
Each reflection, $R_1$, $R_2$ and $R_3$, has $4\cdot 4 = 16$ fixed points: $f^1_{\gb\gg}$, $f^2_{\ga\gg}$ and $f^3_{\ga\gb}$. These singularities are conveniently labeled by $\mu,\nu, \alpha, \beta, \gamma=1,2,3,4= 00, 01,10,11 ;$ {i.e.}\ interpreting them as binary multi--indices $\alpha=\left(\alpha_{1}, \alpha_{2}\right)$ is reserved for the first two--torus, $\beta=\left(\beta_{3}, \beta_{4}\right)$ for the second and $\gamma=\left(\gamma_{5}, \gamma_{6}\right)$ for the third, with the entries take the values $\alpha_{1}, \alpha_{2}, \beta_{3}, \beta_{4}, \gamma_{5}, \gamma_{6}=0,1$. The translation between both conventions read: $\ga = 2\ga_1+\ga_2+1$, $\gb = 2\gb_3+\gb_4+1$ and $\gg = 2\gg_5+\gg_6+1$, respectively.  (The (multi--)indices $\gm,\gn$ are used to label the fixed points in any of the three two--tori in order to write compact expressions.)

Assuming that the tori have unit length, the fixed points may be represented as 
\equ{
f^1_{\gb\gg} = \Big(0, \tfrac {\gb_1+ \gb_2\, i}2, \tfrac{ \gg_1 + \gg_2\, i}2\Big)~, 
\quad 
f^2_{\ga\gg} = \Big(\tfrac {\ga_1+ \ga_2\, i}2, 0, \tfrac{ \gg_1 + \gg_2\, i}2\Big)~, 
\quad 
f^3_{\ga\gb} = \Big(\tfrac {\ga_1+ \ga_2\, i}2, \tfrac{ \gb_1 + \gb_2\, i}2, 0\Big)~.
}
The fixed set of each reflection has the topology of a torus orbifolded by the action of the other orbifold actions which leads to four fixed points on a fixed tori. Hence, in total the $T^6/\Intr_2\times \Intr_2$ orbifold possesses 64 $\Cplx^3/\Intr_2\times \Intr_2$ singularities, 
\equ{
f_{\ga\gb\gg} = \Big(\tfrac {\ga_1+\ga_2\,i}2, \tfrac {\gb_1+ \gb_2\, i}2, \tfrac{ \gg_1 + \gg_2\, i}2\Big)~,
}
coming from every combination of the four fixed points in each of the three complex planes.

\subsection{Geometry of the $\boldsymbol{T^6/\Intr_2\times \Intr_2}$ Resolutions}

The geometry of the resulting resolved orbifolds are characterised by the set of four-cycles (divsors), which are obtained by setting one complex coordinate used in the resolution to zero. There are three classes of divisors~\cite{Lust:2006zh,Blaszczyk:2010db}: 6 inherited divisors $R_{i}:=\left\{u_{i}=0\right\}$ and $R_{i}^{\prime}:=\left\{v_{i}=0\right\}$  that descend from each of the three torus of the orbifold ($u_{i}$ and $v_{i}$, $i=1,2,3$ are the coordinates of the elliptic curves describing the two--dimensional tori that make up $T^6$),  12 ordinary divisors $D_{1, \alpha}:=\left\{z_{1, \alpha}=0\right\}, D_{2, \beta}:=\left\{z_{2, \beta}=0\right\}$, and $D_{3, \gamma}:=$
$\left\{z_{3, \gamma}=0\right\}$ ($z_{i,\mu}$ $i=1,2,3$ are the coordinates of  the covering space)   and finally 48 exceptional divisors $E_{1, \beta \gamma}:=\left\{x_{1, \beta \gamma}=0\right\}, E_{2, \alpha \gamma}:=\left\{x_{2, \alpha \gamma}=0\right\}$, and
$E_{3, \alpha \beta}:=\left\{x_{3, \alpha \beta}=0\right\}$   ($x_{i,\mu\nu}$ are extra coordinates used for the resolution) that appear in the blow--up process.

Not all these divisors are independent; there are a number of linear relations among them, namely:
\equ{ \label{eq:LinearEquivalences} 
\begin{array}{ll}
2 D_{1, \alpha} \sim R_{1}-\sum\limits_{\gamma} E_{2, \alpha \gamma}-\sum\limits_{\beta} E_{3, \alpha \beta}~, & 2 D_{2, \beta} \sim R_{2}-\sum\limits_{\gamma} E_{1, \beta \gamma}-\sum\limits_{\alpha} E_{3, \alpha \beta} \\[2ex] 
2 D_{3, \gamma} \sim R_{3}-\sum\limits_{\beta} E_{1, \beta \gamma}-\sum\limits_{\alpha} E_{2, \alpha \gamma}~, &  
R_{i}^{\prime} \sim R_{i}
\end{array}
}
Here $\sim$ means that these divisors interpreted as $(1,1)$--forms differ by exact forms. 
So in the end  3 $R_{i}$ and  48 $E_{r}$ provide via the Poincar\'e duality a basis of the real cohomology group, {\em i.e.}\ of the $(1,1)$--forms, on the resolved manifold.

\begin{figure}
\begin{center}
\begin{tikzpicture}
 \draw[black, {Triangle[width = 6pt, length = 6pt]}-{Triangle[width = 6pt, length = 6pt]}, line width = 1pt] (2.0, 1.5) -- (3.1, 2.4);
  \draw[black,  line width = 1pt,{Triangle[width = 6pt, length = 6pt]}-{Triangle[width = 6pt, length = 6pt]}] (-1.7, 2.4) -- (-0.6, 1.5); 
  \draw[black,  line width = 1pt,{Triangle[width = 6pt, length = 6pt]}-{Triangle[width = 6pt, length = 6pt]}] (0.9, -1.45) -- (0.9, -2.85);
\foreach \Point/\PointLabel in { (-4.5,3)/D_ {2}}
\draw[fill=black] \Point circle (0.05) node[below left]
{$\PointLabel$};
\foreach \Point/\PointLabel in { (-4.5,4)/E_ {3}}
\draw[fill=black] \Point circle (0.05) node[left]
{$\PointLabel$};
\foreach \Point/\PointLabel in { (-4.5,5)/D_ {1}}
\draw[fill=black] \Point circle (0.05) node[above left]
{$\PointLabel$};
\foreach \Point/\PointLabel in { (-3.5,4)/E_ {2}}
\draw[fill=black] \Point circle (0.05) node[right]
{$\PointLabel$};
\foreach \Point/\PointLabel in { (-2.5,3)/D_ {3}}
\draw[fill=black] \Point circle (0.05) node[below right]
{$\PointLabel$};
\foreach \Point/\PointLabel in { (-3.5,3)/E_ {1}}
\draw[fill=black] \Point circle (0.05) node[below]
{$\PointLabel$};
\draw (-4.5,5) -- (-3.5,3); 
\draw (-3.5,3) -- (-3.5,4); 
\draw (-4.5,4) -- (-3.5,3); 
\draw (-4.5,3) -- (-2.5,3)  -- (-4.5,5) -- (-4.5,3);
\foreach \Point/\PointLabel in { (0,0)/D_ {2}}
\draw[fill=black] \Point circle (0.05) node[below left]
{$\PointLabel$};
\foreach \Point/\PointLabel in { (0,1)/E_ {3}}
\draw[fill=black] \Point circle (0.05) node[left]
{$\PointLabel$};
\foreach \Point/\PointLabel in { (0,2)/D_ {1}}
\draw[fill=black] \Point circle (0.05) node[above left]
{$\PointLabel$};
\foreach \Point/\PointLabel in { (1,1)/E_ {2}}
\draw[fill=black] \Point circle (0.05) node[right]
{$\PointLabel$};
\foreach \Point/\PointLabel in { (2,0)/D_ {3}}
\draw[fill=black] \Point circle (0.05) node[below right]
{$\PointLabel$};
\foreach \Point/\PointLabel in { (1,0)/E_ {1}}
\draw[fill=black] \Point circle (0.05) node[below]
{$\PointLabel$};
\draw (1,0) -- (0,1); 
\draw (1,0) -- (1,1); 
\draw (0,1) -- (1,1); 
\draw (0,0) -- (2,0)  -- (0,2) -- (0,0);
\foreach \Point/\PointLabel in { (4,3)/D_ {2}}
\draw[fill=black] \Point circle (0.05) node[below left]
{$\PointLabel$};
\foreach \Point/\PointLabel in { (4,4)/E_ {3}}
\draw[fill=black] \Point circle (0.05) node[left]
{$\PointLabel$};
\foreach \Point/\PointLabel in { (4,5)/D_ {1}}
\draw[fill=black] \Point circle (0.05) node[above left]
{$\PointLabel$};
\foreach \Point/\PointLabel in { (5,4)/E_ {2}}
\draw[fill=black] \Point circle (0.05) node[right]
{$\PointLabel$};
\foreach \Point/\PointLabel in { (6,3)/D_ {3}}
\draw[fill=black] \Point circle (0.05) node[below right]
{$\PointLabel$};
\foreach \Point/\PointLabel in { (5,3)/E_ {1}}
\draw[fill=black] \Point circle (0.05) node[below]
{$\PointLabel$};
\node[draw,text width=2.675cm] at (1,-0.9) {Triangulation $S$};
\node[draw,text width=2.875cm] at (-3.5,1.9) {Triangulation $E_{1}$};
\node[draw,text width=2.875cm] at (5,1.9) {Triangulation $E_{2}$};
\node[draw,text width=2.875cm] at (1,-5.5) {Triangulation $E_{3}$};
\draw (4,3) -- (5,4); 
\draw (5,3) -- (5,4); 
\draw (4,4) -- (5,4); 
\draw (4,3) -- (6,3)  -- (4,5) -- (4,3);
\foreach \Point/\PointLabel in { (0,-4.6)/D_ {2}}
\draw[fill=black] \Point circle (0.05) node[below left]
{$\PointLabel$};
\foreach \Point/\PointLabel in { (0,-3.6)/E_ {3}}
\draw[fill=black] \Point circle (0.05) node[left]
{$\PointLabel$};
\foreach \Point/\PointLabel in { (0,-2.6)/D_ {1}}
\draw[fill=black] \Point circle (0.05) node[above left]
{$\PointLabel$};
\foreach \Point/\PointLabel in { (1,-3.6)/E_ {2}}
\draw[fill=black] \Point circle (0.05) node[right]
{$\PointLabel$};
\foreach \Point/\PointLabel in { (2,-4.6)/D_ {3}}
\draw[fill=black] \Point circle (0.05) node[below right]
{$\PointLabel$};
\foreach \Point/\PointLabel in { (1,-4.6)/E_ {1}}
\draw[fill=black] \Point circle (0.05) node[below]
{$\PointLabel$};
\draw (1,-4.6) -- (0,-3.6); 
\draw (0,-3.6) -- (2,-4.6); 
\draw (0,-3.6) -- (1,-3.6); 
\draw (0,-4.6) -- (2,-4.6)  -- (0,-2.6) -- (0,-4.6);
\end{tikzpicture}
\end{center}
\caption{ The four different triangulation, the $E_1$--, $E_2$--, $E_3$-- and $S$--triangulation, of the projected toric diagram are given of the resolved $\Cplx^3/\Intr_2\times\Intr_2$. The left--right--arrows indicate the possible flop--transition between different triangulations, which shows that any flop--transition always involves the $S$--triangulation. 
\label{fg:FlopTransitions}}
\end{figure}
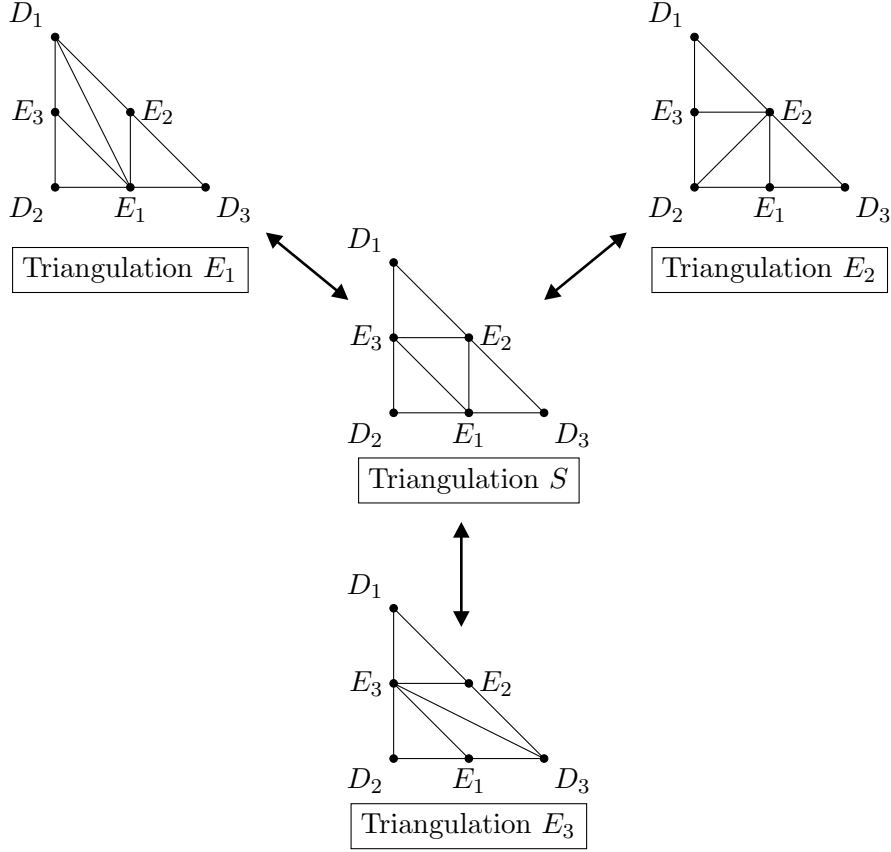

\subsection{Triangulation Dependence and Flop--Transitions} 

To complete the description of the geometry of a resolved orbifold, the intersection numbers of these divisors have to be specified. A major complication to specify the intersection numbers of the resolved $T^6/\Intr_2\times \Intr_2$ orbifold is that there is an indeterminacy, because of the triangulation dependence: each resolved $ \mathbb{C}^{3} / \mathbb{Z}_{2} \times \mathbb{Z}_{2}$ admits four inequivalent resolutions encoded by four different triangulations of the toric diagram of the  $ \mathbb{C}^{3} / \mathbb{Z}_{2} \times \mathbb{Z}_{2}$ singularity. The local projected toric diagrams are given in figure~\ref{fg:FlopTransitions}. There are three triangulations, $E_1$, $E_2$ and $E_3$,  where are all curves, that go through the interior of the projected toric diagram, connect to one of these exceptional divisors. For example in triangulation $E_1$ the curves $E_1E_2$, $E_1E_3$ and $E_1D_1$ all exist. In the final triangulation, dubbed the $S$--triangulation, all the exceptional divisors intersect since the curves $E_1E_2$, $E_2E_3$ and $E_3E_1$ all exist.

The four triangulations of the projected toric diagram given in figure~\ref{fg:FlopTransitions} are related to each other via flop--transitions. From this figure it can be inferred, that the $E_1$, $E_2$ and $E_3$--triangulations are all related via a single flop to the $S$--triangulation. For example, during the flop--transition from the $E_1$--triangulation to the $S$--triangulation, the curve $E_1D_1$ shrinks to zero size and disappears while the curve $E_1E_2$ appears. To go from one $E$--triangulation to another one always has to go through the $S$--triangulation. For example, for the transition from triangulation $E_1$ to $E_2$, first the curve $E_1D_1$ is replaced by the curve $E_2E_3$ to form the $S$--triangulation and after that the curve $E_1E_3$ is replaced by the curve $E_2D_2$ to arrive in the $E_2$--triangulation. This shows that the special role the $S$--triangulation plays in flop--transitions. 

During a flop--transition some curve shrinks to zero size. This means that in this process the effective field theory approximation in the supergravity regime breaks down and stringy corrections could become important. Since, this work only makes use of effective field theory and geometrical methods, the flop--transitions themselves are beyond our description. But the geometries and the spectra on both sides of flops can be determined.

\subsection{Parameterising Triangulations}

\begin{table} 
\[  \renewcommand{\arraystretch}{1.5} 
\begin{array}{|c||cccc|ccc|ccc|}
\hline 
\text{Triangl.} & 
\delta^{E_1}_{\alpha\beta\gamma} &
\delta^{E_2}_{\alpha\beta\gamma} & 
\delta^{E_3}_{\alpha\beta\gamma} &
\delta^{S}_{\alpha\beta\gamma} 
&
\Delta^{1}_{\alpha\beta\gamma} & 
\Delta^{2}_{\alpha\beta\gamma} &
\Delta^{3}_{\alpha\beta\gamma}
&
1-\Delta^{1}_{\alpha\beta\gamma} & 
1-\Delta^{2}_{\alpha\beta\gamma} &
1-\Delta^{3}_{\alpha\beta\gamma}
\\ \hline\hline 
E_1 & 1 & 0 & 0 & 0 & -1 & 1 & 1 & 2 & 0 & 0
\\ 
E_2 & 0 & 1 & 0 & 0 & 1 & -1 & 1 & 0 & 2 & 0
\\ 
E_3 & 0 & 0 & 1 & 0 & 1 & 1 & -1 & 0 & 0 & 2
\\ 
S & 0 & 0 & 0 & 1 & 0 & 0 & 0 & 1 & 1 & 1
\\\hline
\end{array}
\]
\caption{ \label{tb:DeltaTable}
The values of the step functions $\gd^T_{\ga\gb\gg}$ and their variations $\gD^i_{\ga\gb\gg}$, defined in~\eqref{eq:TriangulationStepFunctions} and~\eqref{eq:TriangulationStepVariations}, resp., for the different triangulations are given. 
}
\end{table}

Given that there are four triangulation for each $\Cplx^3/\Intr_2\times \Intr_2$ and  64 $\Intr_2\times \Intr_2$ singularities, this gives a naively total number of $4^{64}$ possibilities (up to some permutation symmetries)~\cite{Blaszczyk:2010db}. As important topological data such as the intersection numbers of the divisors varies for each triangulation, it is particularly useful to develop some formalism to study spectra and the consistency conditions (such as Bianchi identities) for all triangulation choices simultaneously. Next, a formalism will be laid out that is capable of doing just that. 

Define the following four functions: 
\equ{ \label{eq:TriangulationStepFunctions} 
\delta^T_{\alpha\beta\gamma} = 
\begin{cases}
1 & \text{if triangulation }T\text{ is used,} \\[1ex]  
0 & \text{if other triangulation is used,}
\end{cases}
}
of $(\alpha,\beta,\gamma)$ for the four possible triangulations $T =S$, $E_1$, $E_2$ and $E_3$. Since at any of the 64 singularity resolutions one of the four triangulations has to be used, it follows that 
\equ{ \label{eq:ConstraintTriangulationStepFunctions} 
 \delta^{E_1}_{\alpha\beta\gamma}
+\delta^{E_2}_{\alpha\beta\gamma}
+\delta^{E_3}_{\alpha\beta\gamma}
+\delta^{S}_{\alpha\beta\gamma} = 1~. 
} 
Thus, say, $\delta^{S}_{\alpha\beta\gamma}$ is a function of the others. The following combinations of the remaining three independent functions prove particularly useful: 
\equ{ \label{eq:TriangulationStepVariations} 
\begin{array}{rcl}
\Delta^1_{\alpha\beta\gamma} & = & - \delta^{E_1}_{\alpha\beta\gamma}
+ \delta^{E_2}_{\alpha\beta\gamma}
+\delta^{E_3}_{\alpha\beta\gamma}~, 
\\[2ex]
\Delta^2_{\alpha\beta\gamma} & = & \phantom{-} \delta^{E_1}_{\alpha\beta\gamma}
- \delta^{E_2}_{\alpha\beta\gamma}
+\delta^{E_3}_{\alpha\beta\gamma}~, 
\\[2ex] 
\Delta^3_{\alpha\beta\gamma} & = & \phantom{-} \delta^{E_1}_{\alpha\beta\gamma}
+ \delta^{E_2}_{\alpha\beta\gamma}
-\delta^{E_3}_{\alpha\beta\gamma}~. 
\end{array}
}
For example, this means that $\Delta^1_{\alpha\beta\gamma}$ equals $-1$ if singularity $f_{\alpha\beta\gamma}$ is resolved using triangulation $E_1$, $1$ if $E_2$ and $E_3$ and $0$ if $S$. The values that these functions take can be easily read off from the Table~\ref{tb:DeltaTable}. It follows immediately that 
\begin{subequations} 
\equ{
1 
- \Delta^{1}_{\alpha\beta\gamma}
- \Delta^{2}_{\alpha\beta\gamma}
- \Delta^{3}_{\alpha\beta\gamma} 
= \delta^{S}_{\alpha\beta\gamma} ~, 
\qquad 
1 - \Delta^i_{\ga\gb\gg} = 2\, \gd^{E_i}_{\ga\gb\gg} + \gd^{S}_{\ga\gb\gg}~. 
} 
and 
\equ{
\Delta^{2}_{\alpha\beta\gamma} + \Delta^{3}_{\alpha\beta\gamma}  
= 2\, \gd^{E_1}_{\ga\gb\gg}~, 
\qquad 
\Delta^{1}_{\alpha\beta\gamma} + \Delta^{3}_{\alpha\beta\gamma}  
= 2\, \gd^{E_2}_{\ga\gb\gg}~, 
\qquad 
\Delta^{1}_{\alpha\beta\gamma} + \Delta^{2}_{\alpha\beta\gamma}  
= 2\, \gd^{E_3}_{\ga\gb\gg}~. 
}
\end{subequations}

\subsection{Triangulation Dependence of (Self--)Intersections and Chern Classes}

The fundamental (self--)intersection numbers of the basis of divisors read: 
\equ{ \label{eq:SelfIntersections} 
\begin{array}{ll}
R_1E_{1,\beta\gamma}^2 = R_2E_{2,\alpha\gamma}^2 = R_3E_{3,\alpha\beta}^2 = -2~, &
R_1R_2R_3 = 2~, 
\\[2ex] 
E_{1,\beta\gamma}E_{2,\alpha\gamma}^2 = E_{1,\beta\gamma}E_{3,\beta\gamma}^2 = 
-1 + \Delta^1_{\alpha\beta\gamma}~, &
E_{1,\beta\gamma}^3 = \sum\limits_\alpha \big( 1 + \Delta^1_{\alpha\beta\gamma} \big)~, 
\\[2ex] 
E_{2,\alpha\gamma}E_{1,\beta\gamma}^2 = E_{2,\alpha\gamma}E_{3,\beta\gamma}^2 = 
-1 + \Delta^2_{\alpha\beta\gamma}~, &
E_{2,\alpha\gamma}^3 = \sum\limits_\beta \big( 1 + \Delta^2_{\alpha\beta\gamma} \big)~, 
\\[2ex] 
E_{3,\alpha\beta}E_{1,\beta\gamma}^2 = E_{3,\alpha\beta}E_{2,\alpha\gamma}^2 = 
-1 + \Delta^3_{\alpha\beta\gamma}~, &
E_{3,\alpha\beta}^3 = \sum\limits_\gamma \big( 1 + \Delta^3_{\alpha\beta\gamma} \big)~, 
\\[2ex] 
E_{1,\beta\gamma}E_{2,\alpha\gamma}E_{3,\beta\gamma} = 
1 -  \Delta^1_{\alpha\beta\gamma} -  \Delta^2_{\alpha\beta\gamma} - \Delta^3_{\alpha\beta\gamma}~.
\end{array}
} 
and all others are always zero. These (self--)intersection numbers can be partially inferred from the results in ref.~\cite{Blaszczyk:2010db} as follows: as observed in that paper the (partially self--)intersection numbers involving the ordinary divisors $R_i$ are triangulation independent. The (partial self--)intersection numbers involving all three labels $\ga, \gb$ and $\gg$ are fully local, {\em i.e.}\ defined only at the resolution of the single singularity $f_{\ga\gb\gg}$. Thus the intersection numbers for these (partial self--)intersections can be directly read off from Table 4 of ref.~\cite{Blaszczyk:2010db}. (Using the functions $\gD^i_{\ga\gb\gg}$ precisely the local intersection numbers of the four different triangulations of that table are reproduced.) This leaves the cubic self--intersection numbers $E_{1,\gb\gg}^3$, $E_{2,\ga\gg}^3$ and $E_{3,\ga\gb}^3$. But these can be determined using the linear equivalence relations~\eqref{eq:LinearEquivalences}. For example, since the divisors $D_{1,\ga}$, $D_{3,\gg}$ and $E_{2,\ga\gg}$ lie on a straight line in the toric diagram, their intersection vanishes: $D_{1,\ga}E_{2,\ga\gg}D_{3,\gg} =0$. Inserting the linear equivalence relations then leads to the identity
\equ{ 
E_{2,\ga\gg}^3 = - \sum_\gb \Big\{ 
E_{1,\gb\gg} E_{2,\ga\gg}^2 + E_{3,\ga\gb} E_{2,\ga\gg}^2 + 
E_{1,\gb\gg} E_{2,\ga\gg} E_{3,\ga\gb} 
\Big\}
= \sum_\gb \big( 1 + \Delta^2_{\alpha\beta\gamma} \big)~. 
}
This expresses $E_{2,\ga\gg}^3$ in fully local (partial self--)intersection numbers just determined. Inserting those leads to the final expression in this equation. The other two cubic self--intersections are computed in an analogous fashion.

With the fundamental (self--)intersection numbers fixed for any choice of triangulation of all of the 64 resolved $\Intr_2\times\Intr_2$ singularities, all kind of other quantities can be computed. For example, 
the second Chern classes integrated over the basis of divisors can be determined to be given by
\equ{ \label{eq:Integrated2ndChern} 
\begin{array}{ll}
c_2 R_1 = c_2 R_2 = c_2 R_3 = 24~, 
& 
c_2 E_{1,\beta\gamma} = \sum\limits_\alpha \big( 1 - 2  \Delta^1_{\alpha\beta\gamma} \big)~,
\\[2ex] 
c_2 E_{2,\alpha\gamma} = \sum\limits_\beta \big( 1 -2 \Delta^2_{\alpha\beta\gamma}\big)~, 
&
c_2 E_{3,\alpha\beta} = \sum\limits_\gamma \big( 1-2 \Delta^3_{\alpha\beta\gamma}\big)~. 
\end{array}
}

The third Chern class can be evaluated as 
\equ{
c_3 = \dfrac 13 \sum_u (-)^u S_u^3~, 
}
using \eqref{eq:Expansion23Chern} given that the first Chern class vanishes. Since the inherited torus divisors $R_i, R_i'$ have vanishing triple self--intersections, this expression reduces to a sum over all ordinary and exceptional divisors
\equ{
c_3 = \dfrac 13 \sum_{\ga} D_{1,\ga}^3 + \dfrac 13 \sum_{\gb} D_{2,\gb}^3 + \dfrac 13 \sum_{\gg} D_{3,\gg}^3 + 
\dfrac 13 \sum_{\gb.\gg} E_{1,\gb\gg}^3 + \dfrac 13 \sum_{\ga,\gg} E_{2,\ga\gg}^3 + \dfrac 13 \sum_{\ga,\gb} E_{3,\ga\gb}^3~. 
}
The first term can be written as 
\equ{ 
\sum_\ga D_{1,\ga}^3 = 
- \dfrac 18 \sum_{\ga,\gg} E_{2,\ga\gg}^3 - \dfrac 18 \sum_{\ga,\gb} E_{3,\ga\gb}^3 
- \dfrac 38 \sum_{\ga,\gb,\gg} \Big( E_{2,\ga\gg}^2 E_{3,\ga\gb} +   E_{2,\ga\gg} E_{3,\ga\gb}^2 \Big)~, 
}
using that there are no non--vanishing intersections of $R_1$ with $E_{2,\ga\gg}$ or $E_{3,\ga\gb}$. Adding similar expressions involving $D_{2,\gb}$ and $D_{3,\gg}$, one can show that 
\equ{
 c_3 = - \dfrac 18 \sum_{\ga,\gb,\gg} \Big\{ 
E_{1,\gb\gg} \big( E_{2,\ga\gg}^2 + E_{3,\ga\gb}^2 \big) + 
E_{2,\ga\gg} \big( E_{1,\gb\gg}^2 + E_{3,\ga\gb}^2 \big) + 
E_{3,\ga\gb} \big( E_{1,\gb\gg}^2 + E_{2,\ga\gg}^2 \big)
\Big\} 
\non \\
+ \dfrac 14 \sum_{\gb.\gg} E_{1,\gb\gg}^3 
+ \dfrac 14 \sum_{\ga,\gg} E_{2,\ga\gg}^3 
+ \dfrac 14 \sum_{\ga,\gb} E_{3,\ga\gb}^3~. 
}
Finally, inserting the triangulation dependent intersection numbers~\eqref{eq:SelfIntersections}, gives 
\equ{
c_3 = \dfrac 14 \sum_{i, \ga,\gb,\gg} \Big( 1 + \gD_{\ga\gb\gg}^i\Big) 
-\dfrac 14 \sum_{i, \ga,\gb,\gg} \Big( -1 + \gD_{\ga\gb\gg}^i\Big) = 96~.  
}
Note, in particular, that all the triangulation dependence in the form of the functions $\gD^i_{\ga\gb\gg}$ drops out and the final result equals the well--known Euler number 96.

\subsection{Line Bundle Backgrounds}

The line bundle backgrounds considered in this paper only have flux supported on the exceptional cycles: 
\equ{
\frac{\cF}{2\pi} = 
\sum_{i,\gm,\gn} E_{i,\gm\gn}\, \mathsf{H}_{i,\gm\gn}~, 
\qquad 
\mathsf{H}_{i,\gm\gn} = \sum_I \cV_{i,\gm\gn}^I\, \mathsf{H}_I~. 
}
Here the Cartan generators $\mathsf{H}_I$ are anti--Hermitian and therefore so is the field strength $\cF$. The entries of the line bundle vectors $\cV_{i,\gm\gn}$ are subject to flux quantisation conditions which are triangulation dependent: 
\equ{ \label{eq:FluxQuantisation} 
\int_C \frac{\cF}{2\pi} = L^I \, \mathsf{H}_I~,
\qquad 
L \cong 0~,
}
where $\cong$ means equal up to $E_8\times E_8$ lattice vectors, for any $C$ inside the resolved orbifold. The resulting conditions for any choice of triangulation are listed in Table~\ref{tb:FluxQuantisation}.

\begin{table}[h!t]
\[
\renewcommand{\arraystretch}{1.6} 
\arry{|c|c||c|c|}{
\hline 
\multicolumn{4}{|c|}{\text{Flux quantisation conditions for arbitrary triangulations}}
\\ \hline \hline  
R_1E_{1,\gb\gg} & 
2\, \cV_{1,\gb\gg}  \cong 0 
&
D_{1,\ga} E_{1,\gb\gg} & 
 \big( \cV_{1,\beta\gamma}-\cV_{2,\alpha\gamma}-\cV_{3,\alpha\beta} \big) \gd^{E_1}_{\ga\gb\gg} \cong 0
\\ \hline 
R_2E_{2,\ga\gg} & 
2\, \cV_{2,\ga\gg}  \cong 0 
&
D_{2,\gb} E_{2,\ga\gg} & 
 \big( \cV_{2,\alpha\gamma}-\cV_{1,\beta\gamma}-\cV_{3,\alpha\beta} \big) \gd^{E_2}_{\ga\gb\gg}  \cong 0
\\ \hline 
R_3E_{3,\ga\gb} & 
2\, \cV_{3,\ga\gb}  \cong 0 
&
D_{3,\gg} E_{3,\ga\gb} & 
 \big( \cV_{3,\alpha\beta} - \cV_{1,\beta\gamma}-\cV_{2,\alpha\gamma} \big) \gd^{E_3}_{\ga\gb\gg}  \cong 0
\\ \hline\hline 
R_{1}D_{2,\gb}  & 
 - \sum\limits_{\gg} \cV_{1,\gb\gg} \cong 0
& 
D_{1,\ga} E_{2,\ga\gg} & 
 - \sum\limits_\gb \Big\lbrace 
\cV_{3,\ga\gb} + \big( \cV_{1,\beta\gamma}-\cV_{2,\alpha\gamma}-\cV_{3,\alpha\beta} \big) \gd^{E_1}_{\ga\gb\gg}
\Big\rbrace \cong 0
\\ \hline 
R_{1}D_{3,\gg}  & 
 - \sum\limits_{\gb} \cV_{1,\gb\gg} \cong 0
& 
D_{1,\ga} E_{3,\ga\gb} & 
 - \sum\limits_\gg \Big\lbrace 
\cV_{2,\ga\gg} + \big( \cV_{1,\beta\gamma}-\cV_{2,\alpha\gamma}-\cV_{3,\alpha\beta} \big) \gd^{E_1}_{\ga\gb\gg}
\Big\rbrace \cong 0
\\ \hline 
R_{2}D_{1,\ga}  & 
 - \sum\limits_{\gg} \cV_{2,\ga\gg} \cong 0
& 
D_{2,\gb} E_{1,\gb\gg} & 
 - \sum\limits_\ga \Big\lbrace 
\cV_{3,\ga\gb} + \big( \cV_{2,\alpha\gamma}-\cV_{1,\beta\gamma}-\cV_{3,\alpha\beta} \big) \gd^{E_2}_{\ga\gb\gg}
\Big\rbrace \cong 0
\\ \hline 
R_{2}D_{3,\gg}  & 
 - \sum\limits_{\ga} \cV_{2,\ga\gg} \cong 0
& 
D_{2,\gb} E_{3,\ga\gb} & 
 - \sum\limits_\gg \Big\lbrace 
\cV_{1,\gb\gg} + \big( \cV_{2,\alpha\gamma}-\cV_{1,\beta\gamma}-\cV_{3,\alpha\beta} \big) \gd^{E_2}_{\ga\gb\gg}
\Big\rbrace \cong 0
\\ \hline 
R_{3}D_{1,\ga}  & 
 - \sum\limits_{\gb} \cV_{3,\ga\gb} \cong 0
& 
D_{3,\gg} E_{1,\gb\gg} & 
 - \sum\limits_\ga \Big\lbrace 
\cV_{2,\ga\gg} + \big( \cV_{3,\alpha\beta}-\cV_{1,\beta\gamma}-\cV_{2,\alpha\gamma} \big) \gd^{E_3}_{\ga\gb\gg} 
\Big\rbrace\cong 0
\\ \hline 
R_{3}D_{2,\gb}  & 
 - \sum\limits_{\ga} \cV_{3,\ga\gb} \cong 0
& 
D_{3,\gg} E_{2,\ga\gg} & 
 - \sum\limits_\gb \Big\lbrace 
\cV_{1,\gb\gg} + \big(\cV_{3,\alpha\beta} - \cV_{1,\beta\gamma}-\cV_{2,\alpha\gamma}\big) \gd^{E_3}_{\ga\gb\gg}
\Big\rbrace \cong 0
\\ \hline\hline 
E_{1,\gb\gg}E_{2,\ga\gg} & 
\multicolumn{3}{c|}{ 
2\,\cV_{2,\ga\gg}\, \gd^{E_1}_{\ga\gb\gg}  + 2\,\cV_{1,\gb\gg}\, \gd^{E_2}_{\ga\gb\gg} + 
 \big( \cV_{1,\beta\gamma}+\cV_{2,\alpha\gamma}-\cV_{3,\alpha\beta} \big) \gd^{S}_{\ga\gb\gg} \cong 0 }
\\\hline
E_{1,\gb\gg}E_{3,\ga\gb} & 
\multicolumn{3}{c|}{ 
2\,\cV_{3,\ga\gb}\, \gd^{E_1}_{\ga\gb\gg}  + 2\,\cV_{1,\gb\gg}\, \gd^{E_3}_{\ga\gb\gg} + 
 \big( \cV_{1,\beta\gamma}+\cV_{3,\alpha\beta}-\cV_{2,\alpha\gamma} \big) \gd^{S}_{\ga\gb\gg} \cong 0 }
\\\hline
E_{2,\ga\gg}E_{3,\ga\gb} & 
\multicolumn{3}{c|}{ 
2\,\cV_{3,\ga\gb}\, \gd^{E_2}_{\ga\gb\gg}  + 2\,\cV_{2,\ga\gg}\, \gd^{E_3}_{\ga\gb\gg} + 
 \big( \cV_{2,\alpha\gamma}+\cV_{3,\alpha\beta} - \cV_{1,\beta\gamma} \big) \gd^{S}_{\ga\gb\gg} \cong 0 }
\\\hline
}
\]
\renewcommand{\arraystretch}{1} 
\caption{The flux quantisation conditions on the line bundle vectors $\cV_{i,\gm\gn}$ the resolved orbifold $X$ using arbitrary triangulation at the 64 $\Cplx^3/\ztwo$ resolutions.  
\label{tb:FluxQuantisation}}
\end{table}

\subsection{General Bianchi Identities}

Consistency of the effective field theory description demands that the integrated Bianchi identity
\equ{ \label{eq:IntegratedBianchiIdentities} 
\int_D \Big\{ \tr \cF^2 - \tr \cR^2\Big\} = 0
}
over any divisor $D$ vanishes. Here $\cR$ denotes the anti--Hermitian curvature two--form. (When non--perturbative contributions of heterotic five--branes are taken into account this condition can be weakened somewhat~\cite{Honecker:2006dt}.) By considering the basis of divisors spanned by the ordinary divisors $R_i$ and the exceptional divisors $E_{1,\gb\gg}$, $E_{2,\ga\gg}$ and $E_{3,\ga\gb}$ the complete set of integrated Bianchi identities is obtained.

The three Bianchi identities on the three ordinary divisors, $R_1$, $R_2$ and $R_3$ are the ones one expects on K3 surfaces: 
\begin{subequations} 
\begin{equation} 
\sum_{\beta,\gamma} \cV_{1,\beta\gamma}^2 = 24~, 
\quad 
\sum_{\alpha,\gamma} \cV_{2,\alpha\gamma}^2 = 24~, 
\quad 
\sum_{\alpha,\beta} \cV_{3,\alpha\beta}^2 = 24~,  
\end{equation} 
and do not depend on the triangulations chosen. In contrast the Bianchi identities on the exceptional divisors are very sensitive to the triangulations used in the local resolutions. The sixteen Bianchi identities on $E_{1,\beta\gamma}$ take the form
\equ{
\sum\limits_\alpha \Big[ 
(1+\Delta^1_{\alpha\beta\gamma}) \cV^2_{1,\beta\gamma} + 
(-1+\Delta^1_{\alpha\beta\gamma}) (\cV^2_{2,\alpha\gamma} + \cV^2_{3,\alpha\beta})
+2(1-\Delta^1_{\alpha\beta\gamma}-\Delta^2_{\alpha\beta\gamma}-\Delta^3_{\alpha\beta\gamma}) \cV_{2,\alpha\gamma}\cdot \cV_{3,\alpha\beta} 
 \non \\
+2(-1+\Delta^2_{\alpha\beta\gamma}) \cV_{1,\beta\gamma}\cdot \cV_{2,\alpha\gamma}  
+2(-1+\Delta^3_{\alpha\beta\gamma}) \cV_{1,\beta\gamma}\cdot \cV_{3,\alpha\beta} 
 \Big] = 
 \sum\limits_\alpha \Big[ -2 + 4\, \Delta^1_{\alpha\beta\gamma} \Big]~. 
} 
The sixteen Bianchi identities on $E_{2,\alpha\gamma}$ take the form
\equ{
\sum\limits_\beta \Big[ 
(1+\Delta^2_{\alpha\beta\gamma}) \cV^2_{2,\alpha\gamma} + 
(-1+\Delta^2_{\alpha\beta\gamma}) (\cV^2_{1,\beta\gamma} + \cV^2_{3,\alpha\gamma})
+2(1-\Delta^1_{\alpha\beta\gamma}-\Delta^2_{\alpha\beta\gamma}-\Delta^3_{\alpha\beta\gamma}) \cV_{1,\beta\gamma}\cdot \cV_{3,\alpha\beta} 
 \non \\
+2(-1+\Delta^1_{\alpha\beta\gamma}) \cV_{2,\alpha\gamma}\cdot \cV_{1,\beta\gamma}  
+2(-1+\Delta^3_{\alpha\beta\gamma}) \cV_{2,\alpha\gamma}\cdot \cV_{2,\alpha\gamma} 
 \Big] = 
 \sum\limits_\beta \Big[ -2 + 4\, \Delta^2_{\alpha\beta\gamma} \Big]~. 
} 
And finally, the sixteen Bianchi identities on $E_{3,\alpha\beta}$ take the form
\equ{
\sum\limits_\gamma \Big[ 
(1+\Delta^3_{\alpha\beta\gamma}) \cV^2_{3,\alpha\beta} + 
(-1+\Delta^3_{\alpha\beta\gamma}) (\cV^2_{1,\beta\gamma} + \cV^2_{2,\alpha\gamma})
+2(1-\Delta^1_{\alpha\beta\gamma}-\Delta^2_{\alpha\beta\gamma}-\Delta^3_{\alpha\beta\gamma}) \cV_{1,\beta\gamma}\cdot \cV_{2,\alpha\gamma} 
\non  \\
+2(-1+\Delta^1_{\alpha\beta\gamma}) \cV_{3,\alpha\beta}\cdot \cV_{1,\beta\gamma}  
+2(-1+\Delta^2_{\alpha\beta\gamma}) \cV_{3,\alpha\beta}\cdot \cV_{2,\alpha\gamma} 
 \Big] = 
 \sum\limits_\gamma \Big[ -2 + 4\, \Delta^3_{\alpha\beta\gamma} \Big]~. 
} 
\end{subequations} 

\subsection{Multiplicity Operators}

A convenient tool to compute the chiral spectrum on a resolution with a line bundle background is the multiplicity operator $\mathsf{N}$. It reads~\cite{Nibbelink:2007rd,Nibbelink:2007pn}: 
\equ{
\mathsf{N} = 
\int_X \Big\{ 
\dfrac 1{6} \Big( \dfrac{\cF}{2\pi}\Big)^2 
- \dfrac 1{24} \Big(\dfrac{\cR}{2\pi}\Big)^2
\Big\}  \dfrac{\cF}{2\pi}
}
and may be thought of as a representation dependent index. Hence, on all states it should be integral provided that the fundamental consistency conditions, flux quantisation and the integrated Bianchi identities, are fulfilled.

On the $T^6/\Intr_2\times\Intr_2$ resolutions the multiplicity operator can be evaluated to be equal to:
%
\equ{ \label{MultiplicityOperator} 
\begin{array}{rl}
\mathsf{N} = \sum\limits_{\alpha,\beta,\gamma}\Big[ 
& \mathsf{H}_{1,\beta\gamma} \Big\{
\tfrac 13(\mathsf{H}_{1,\beta\gamma}^2 - \tfrac 14)
-\big(1-\Delta^1_{\alpha\beta\gamma}  \big) 
\Big(\tfrac 16 (\mathsf{H}_{1,\beta\gamma}^2 -1) + \tfrac 12 (\mathsf{H}_{2,\alpha\gamma} - \mathsf{H}_{3,\alpha\beta})^2  \Big) \Big\}
 \\[2ex] 
& \!\!\!\!\!+\mathsf{H}_{2,\alpha\gamma} \Big\{
\tfrac 13 (\mathsf{H}_{2,\alpha\gamma}^2 - \tfrac 14) 
-\big(1-\Delta^2_{\alpha\beta\gamma}  \big) 
\Big( \tfrac 16 (\mathsf{H}_{2,\alpha\gamma}^2 -1) + \tfrac 12 (\mathsf{H}_{1,\beta\gamma} - \mathsf{H}_{3,\alpha\beta})^2  \Big) \Big\}
 \\[2ex] 
& \!\!\!\!\!+\mathsf{H}_{3,\alpha\beta} \Big\{
 \tfrac 13 (\mathsf{H}_{3,\alpha\beta}^2 - \tfrac 14)
-\big(1-\Delta^3_{\alpha\beta\gamma} \big) 
\Big( \tfrac 16 (\mathsf{H}_{3,\alpha\beta}^2 -1) + \tfrac 12 (\mathsf{H}_{1,\beta\gamma} - \mathsf{H}_{2,\alpha\gamma})^2  \Big) \Big\}
\\[2ex] 
& \!\!\!\!\!
- 2\mathsf{H}_{1,\beta\gamma} \mathsf{H}_{2,\alpha\gamma} \mathsf{H}_{3,\alpha\beta}~~
\Big]~.
\end{array}
}
The triangulation dependance is isolated to the second terms on the first three lines of this expression. From Table~\ref{tb:DeltaTable} it may be inferred that only the terms in the first line are switched on (with a multiplicative factor of 2) if triangulation $E_1$ is chosen, the second for $E_2$ and the third for $E_3$; all three are switched on (with a factor 1) for triangulation $S$.

Using the constraint~\eqref{eq:ConstraintTriangulationStepFunctions} another representation of this operator can be obtained 
\equ{ \label{eq:TriangulationDependentMultiplicities} 
\mathsf{N} = \sum_{\ga,\gb,\gg} \Big[ 
\gd^{E_1}_{\ga\gb\gg}\, \mathsf{N}^{E_1}_{\ga\gb\gg} + 
\gd^{E_2}_{\ga\gb\gg}\, \mathsf{N}^{E_2}_{\ga\gb\gg} + 
\gd^{E_3}_{\ga\gb\gg}\, \mathsf{N}^{E_3}_{\ga\gb\gg} + 
\gd^{S}_{\ga\gb\gg}\, \mathsf{N}^{S}_{\ga\gb\gg}
\Big]~, 
} 
where 
\begin{subequations} \label{eq:LocalMultiplicities} 
\equa{
 \mathsf{N}^{E_1}_{\ga\gb\gg} &= 
\tfrac 14\, \mathsf{H}_{1,\gb\gg} 
+ \tfrac 1{12}\, \mathsf{H}_{2,\ga\gg} \big( 4 \mathsf{H}_{2,\ga\gg}^2-1\big) 
+ \tfrac 1{12}\, \mathsf{H}_{3,\ga\gb}\big( 4 \mathsf{H}_{3,\ga\gb}^2-1\big) 
- \mathsf{H}_{1,\gb\gg} \big( \mathsf{H}_{2,\ga\gg}^2 + \mathsf{H}_{3,\ga\gb}^2 \big)~, 
\\[1ex] 
 \mathsf{N}^{E_2}_{\ga\gb\gg} &= 
\tfrac 14\, \mathsf{H}_{2,\ga\gg} 
+ \tfrac 1{12}\, \mathsf{H}_{1,\gb\gg} \big( 4 \mathsf{H}_{1,\gb\gg}^2-1\big) 
+ \tfrac 1{12}\, \mathsf{H}_{3,\ga\gb} \big( 4 \mathsf{H}_{3,\ga\gb}^2-1\big) 
- \mathsf{H}_{2,\ga\gg} \big( \mathsf{H}_{1,\gb\gg}^2 + \mathsf{H}_{3,\ga\gb}^2 \big)~, 
\\[1ex] 
 \mathsf{N}^{E_3}_{\ga\gb\gg} &= 
\tfrac 14\, \mathsf{H}_{3,\ga\gb} 
+\tfrac 1{12}\, \mathsf{H}_{1,\ga\gb} \big( 4 \mathsf{H}_{1,\ga\gb}^2-1\big) 
+ \tfrac 1{12}\, \mathsf{H}_{2,\ga\gg} \big( 4 \mathsf{H}_{2,\ga\gg}^2-1\big) 
- \mathsf{H}_{3,\ga\gb} \big( \mathsf{H}_{1,\gb\gg}^2 + \mathsf{H}_{2,\ga\gg}^2 \big)~, 
\\[1ex] 
 \mathsf{N}^{S}_{\ga\gb\gg} &= \non 
\tfrac 1{12}\, \mathsf{H}_{1,\gb\gg}\big( 2\mathsf{H}_{1,\gb\gg}^2 +1) 
+\tfrac 1{12}\, \mathsf{H}_{2,\ga\gg}\big( 2\mathsf{H}_{2,\ga\gg}^2 +1) 
+\tfrac 1{12}\, \mathsf{H}_{3,\ga\gb}\big( 2\mathsf{H}_{3,\ga\gb}^2 +1) 
+\mathsf{H}_{1,\gb\gg} \mathsf{H}_{2,\ga\gg} \mathsf{H}_{3,\ga\gb} 
\\[1ex] 
&\phantom{=} 
-\tfrac 12\, \mathsf{H}_{1,\gb\gg}\big(\mathsf{H}_{2,\ga\gg}^2 + \mathsf{H}_{3,\ga\gb}^2\big)
-\tfrac 12\, \mathsf{H}_{2,\ga\gg} \big( \mathsf{H}_{1,\gb\gg}^2 + \mathsf{H}_{3,\ga\gb}^2\big)
 -\tfrac 12\, \mathsf{H}_{3,\ga\gb} \big( \mathsf{H}_{1,\gb\gg}^2+\mathsf{H}_{2,\ga\gg}^2\big)~. 
}
\end{subequations} 
These operators can be thought of as the local resolution multiplicities at the resolved singularity $(\ga,\gb,\gg)$ using one of the four triangulations. In particular, when taking the same triangulation at all fixed points, the expressions (56) and (58) of ref.~\cite{Blaszczyk:2010db} are obtained from \eqref{eq:TriangulationDependentMultiplicities}. In general, \eqref{eq:TriangulationDependentMultiplicities} implies that the spectrum in any triangulation can be determined from the local resolution operators~\eqref{eq:LocalMultiplicities} times the functions that indicate which triangulation has been used at each of the 64 $\Cplx^3/\Intr_2\times\Intr_2$ resolved singularities. It should be emphasised that these local multiplicity operators $\mathsf{N}^T_{\ga\gb\gg}$ for a given triangulation $T$ are not necessarily all integral; only their combination in~\eqref{eq:TriangulationDependentMultiplicities} in general is.

\subsection{Jumping Spectra due to Flop--Transitions}
\label{sc:FlopTransitionSpectraJumps}

For a flop--transition to be possible it is necessary that all fundamental consistency conditions, like flux quantisation and the Bianchi identities, have to hold for both triangulation choices before and after the flop. Note that this implies, that if at some resolved singularity $f_{\ga\gb\gg}$ some of these fundamental consistency conditions are not fulfilled for triangulation $S$, then no flop--transitions can occur and resolution is {\em frozen} in one of the three triangulations $E_1, E_2$ or $E_3$. Moreover, if at all resolved $\Intr_2\times\Intr_2$ singularities triangulation $S$ is not admissible, no flop--transition is possible at all!

Assuming that at a resolved singularity $f_{\ga\gb\gg}$ a flop--transition can occur between triangulations $S$ to $E_i$, the difference multiplicity 
\equ{ \label{eq:DifferenceMultiplicities}
\gD \mathsf{N}^{i}_{\ga\gb\gg} = \mathsf{N}^{E_i}_{\ga\gb\gg} - \mathsf{N}^{S}_{\ga\gb\gg}
}
measures the jump in the spectra when the flop--transition goes from triangulation $S$ to $E_i$; $-\gD \mathsf{N}^{i}_{\ga\gb\gg}$ the spectra jump in the opposite direction. This difference multiplicity operator has to be integral because the multiplicity operator~\eqref{MultiplicityOperator} before and after the flop--transition is integral by an index theorem (since the fundamental consistency conditions are assumed to be fulfilled) and this operator is simply the difference of the spectra in the two cases.

\subsection{Volumes and the DUY equations}

\begin{table}[h!t]
\[
\renewcommand{\arraystretch}{1.6} 
\arry{|c|c||c|c|}{
\hline 
\multicolumn{4}{|c|}{\text{Curves}}
\\ \hline\hline 
%
%

R_1R_2 & 
2\, a_3 
&
D_{1,\ga} E_{1,\gb\gg} & 
 \big( b_{1,\beta\gamma}-b_{2,\alpha\gamma}-b_{3,\alpha\beta} \big) \gd^{E_1}_{\ga\gb\gg}
\\ \hline 
R_1E_{1,\gb\gg} & 
2\, b_{1,\gb\gg}  
& 
D_{1,\ga} E_{2,\ga\gg} & 
a_2 - \sum\limits_\gb \Big\lbrace 
b_{3,\ga\gb} + \big( b_{1,\beta\gamma}-b_{2,\alpha\gamma}-b_{3,\alpha\beta} \big) \gd^{E_1}_{\ga\gb\gg}
\Big\rbrace
\\ \hline 
R_{1}D_{2,\gb}  & 
a_3 - \sum\limits_{\gg} b_{1,\gb\gg} 
&
E_{1,\gb\gg}E_{2,\ga\gg} & 
2\,b_{2,\ga\gg}\, \gd^{E_1}_{\ga\gb\gg}  + 2\,b_{1,\gb\gg}\, \gd^{E_2}_{\ga\gb\gg} + 
 \big( b_{1,\beta\gamma}+b_{2,\alpha\gamma}-b_{3,\alpha\beta} \big) \gd^{S}_{\ga\gb\gg}
%
%
\\\hline\hline 
\multicolumn{4}{|c|}{\text{Divisors}}
\\ \hline\hline 
R_1 & 
\multicolumn{3}{c|}{
2\,a_2 a_3 - \sum\limits_{\gb,\gg} b_{1,\gb\gg}^2 
}
 \\ \hline 
D_{1,\alpha} & \multicolumn{3}{c|}{
 a_2 a_3 - \sum\limits_\gamma a_2 b_{2,\alpha\gamma} - \sum\limits_\beta a_3 b_{3,\alpha\beta} 
 + \sum\limits_{\beta,\gamma} 
 \big( 1 - \gd^{E_1}_{\ga\gb\gg} \big)  b_{2,\alpha\gamma} b_{3,\alpha\beta} 
 } 
 \\
& \multicolumn{3}{c|}{
+ \sum\limits_{\beta,\gamma}  \gd^{E_1}_{\ga\gb\gg} \Big\lbrace
 b_{1,\gb\gg} \big(b_{2,\ga\gg} + b_{3,\ga\gb}\big) 
 - \frac 12 \big( b_{1,\gb\gg}^2  + b_{2,\ga\gg}^2  + b_{3,\ga\gb}^2 \big) 
\Big\rbrace 
}
\\ \hline 
E_{1,\beta\gamma} & 
 \multicolumn{3}{c|}{
2\,a_1 b_{1,\beta\gamma}    
 + \sum\limits_\alpha \Big\lbrace
\frac 12 \big( 1 + \gD^1_{\ga\gb\gg}\big) b_{1,\beta\gamma}^2
+\big(1 -\gD^1_{\ga\gb\gg} - \gD^2_{\ga\gb\gg} - \gD^3_{\ga\gb\gg}\big) b_{2,\alpha\gamma} b_{3,\alpha\beta} 
}
\\
 & 
 \multicolumn{3}{c|}{
- \frac 12 \big( 1 - \gD^1_{\ga\gb\gg}\big) \big( b_{2,\alpha\gamma}^2 + b_{3,\alpha\beta}^2 \big) 
- \big( 1 - \gD^2_{\ga\gb\gg}\big) b_{1,\beta\gamma} b_{2,\ga\gg} 
- \big( 1 - \gD^3_{\ga\gb\gg}\big) b_{1,\beta\gamma} b_{3,\ga\gb} 
\Big\rbrace
}
\\\hline\hline
\multicolumn{4}{|c|}{\text{Full manifold}}
\\ \hline\hline 
%
%
X & 
\multicolumn{3}{c|}{
2\, a_1 a_2 a_3 - \sum\limits_{\beta,\gamma} a_1 b_{1,\beta\gamma}^2 - \sum\limits_{\alpha,\gamma} a_2 b_{2,\alpha\gamma}^2 - \sum\limits_{\alpha,\beta} a_3 b_{3,\alpha\beta}^2
- \sum\limits_{\alpha,\beta,\gamma} \Big\lbrace 
 \frac{1}{2} \big(\gD^1_{\ga\gb\gg}-1\big)  
 b_{1,\beta\gamma}\big(b_{2,\alpha\gamma}^2 + b_{3,\alpha\beta}^2 \big) 
}
\\ & \multicolumn{3}{c|}{
  +\frac{1}{2} \big(\gD^2_{\ga\gb\gg}-1\big)  
 b_{2,\alpha\gamma}\big(b_{1,\beta\gamma}^2 + b_{3,\alpha\beta}^2 \big) 
 +\frac{1}{2} \big(\gD^3_{\ga\gb\gg}-1\big)  
 b_{3,\alpha\beta}\big(b_{1,\beta\gamma}^2 + b_{2,\alpha\gamma}^2 \big)
 + \frac{1}{6}\big( 1+\gD^1_{\ga\gb\gg}\big) b_{1,\gb\gg}^3 
 }
\\ & \multicolumn{3}{c|}{ 
+\frac{1}{6}\big( 1+\gD^2_{\ga\gb\gg}\big) b_{2,\ga\gg}^3 
+\frac{1}{6}\big( 1+\gD^3_{\ga\gb\gg}\big) b_{3,\ga\gg}^3
+\big(1 -\gD^1_{\ga\gb\gg} - \gD^2_{\ga\gb\gg} - \gD^3_{\ga\gb\gg}\big) b_{1,\beta\gamma} b_{2,\alpha\gamma} b_{3,\alpha\beta}
\Big\rbrace 
} 
\\\hline 
}
\]
\renewcommand{\arraystretch}{1} 
\caption{Volume of a collection of possibly existing curves, divisors and the resolved orbifold $X$ as a whole using arbitrary triangulation at the 64 $\Cplx^3/\ztwo$ resolutions. Similar expression of the other curves and divisors can be obtained by permutations. 
\label{tb:ResolutionVolumes}}
\end{table}

Using the (self--)intersections \eqref{eq:SelfIntersections} various volumes can be computed using the K\"ahler form 
\equ{
J = \sum_i a_i R_i - \sum_r b_r E_r~,
}
involving the K\"ahler parameters $a_i$ and $b_r$. The volumes of a curve $C$, a divisor $D$ and the orbifold resolution $X$  are given by 
\equ{
\text{Vol}(C) = \int_C J~, 
\qquad
\text{Vol}(D) = \int_D \frac 12\,J^2~, 
\qquad
\text{Vol}(X) = \int_X \frac 1{3!}\, J^3~, 
}
respectively. The resulting expressions for any choice of triangulation are given in Table~\ref{tb:ResolutionVolumes}.

The volumes of the divisors are constrained by the DUY equations~\cite{Donaldson:1985,Uhlenbeck:1986}. The one--loop corrections to these equations are given by~\cite{Blumenhagen:2005ga,Nibbelink:2009sp}
\equ{
\int \dfrac 12 J^2\, \dfrac{\cF}{2\pi} = \dfrac {e^{2\gf}}{16\pi} \int 
\Big\{ \tr \Big( \dfrac {\cF'}{2\gp} \Big)^2 - \dfrac 12 \tr \Big(\dfrac{\cR}{2\gp} \Big)^2 \Big\} \dfrac{\cF'}{2\gp} + ('\ra'')~,  
}
where $\cF'$ and $\cF''$ denote the Abelian gauge fluxes in the first and second factor of the $E_8\times E_8$ group, respectively, so that $\cF = \cF' + \cF''$. This equation thus links the K\"ahler moduli $a_i, b_r$ and the dilaton $\gf$ in general.

If the gauge background is embedded in just a single, say first $E_8$, or if one considers the heterotic $SO(32)$ theory instead, this equation may be rewritten as 
\equ{ 
\int \dfrac 12 J^2\, \dfrac{\cF}{2\pi} = \dfrac {e^{2\gf}}{32\pi} \int \tr \Big(\dfrac{\cR}{2\gp} \Big)^2  \dfrac{\cF}{2\gp}
= 
- \dfrac {e^{2\gf}}{16\pi} \int c_2 \dfrac{\cF}{2\gp}~,  
}
as $\cF= \cF'$ and $\cF'' = 0$ using the integrate Bianchi identities~\eqref{eq:IntegratedBianchiIdentities}. Inserting the expansion for the gauge flux in terms of the exceptional divisors $E_r$ and using the integrated second Chern classes~\eqref{eq:Integrated2ndChern}, leads to 
\equ{ 
\sum_{\gb,\gg} \Big\{ \text{Vol}(E_{1,\gb\gg}) - \dfrac{e^{2\gf}}{16\gp} \sum_\ga (1 - 2\gD^1_{\ga\gb\gg}) \Big\} \cV^I_{1,\gb\gg}  + 
\sum_{\ga,\gg} \Big\{ \text{Vol}(E_{2,\ga\gg}) - \dfrac{e^{2\gf}}{16\gp} \sum_\gb (1 - 2\gD^2_{\ga\gb\gg}) \Big\} \cV^I_{2,\ga\gg} + 
\non \\ \label{eq:SingleE8DUYs} 
+ 
\sum_{\ga,\gb} \Big\{ \text{Vol}(E_{3,\ga\gb}) - \dfrac{e^{2\gf}}{16\gp} \sum_\gg (1 - 2\gD^3_{\ga\gb\gg}) \Big\} \cV^I_{3,\ga\gb}  = 0~.  
}

If the gauge background lie in both $E_8$ factors simultaneously, then the $\tr \cR^2$ term can be eliminated using the Bianchi identities~\eqref{eq:IntegratedBianchiIdentities} instead. Moreover, since both $E_8$ factors are independent, the DUY equation may be split into two equations; one for each $E_8$ factor: 
\equ{
\arry{rcl}{ \dsp 
\int \dfrac 12 J^2\, \dfrac{\cF'}{2\pi} &=& \dsp 
\phantom{-} \dfrac {e^{2\gf}}{32\pi} \int 
\Big\{ \tr \Big( \dfrac {\cF'}{2\gp} \Big)^2 - \tr \Big(\dfrac{\cF''}{2\gp} \Big)^2 \Big\} \dfrac{\cF'}{2\gp}~,  
\\[2ex] 
 \dsp 
\int \dfrac 12 J^2\, \dfrac{\cF''}{2\pi} &=& \dsp 
-\dfrac {e^{2\gf}}{32\pi} \int 
\Big\{ \tr \Big( \dfrac {\cF'}{2\gp} \Big)^2 - \tr \Big(\dfrac{\cF''}{2\gp} \Big)^2 \Big\} \dfrac{\cF''}{2\gp}~. 
}
}
Notice the relative sign difference between the otherwise very similar expressions in both $E_8$'s. Evaluating these expressions further by inserting the intersection numbers~\eqref{eq:SelfIntersections} leads to rather lengthy and not very illuminating expressions. For this reason we refrain from stating them here.

\section{Triangulation Independence}
\label{sc:TriangulationIndependence} 

The results obtained in the previous section hold for any particular choice of the triangulation of each of the 64 resolved $\Cplx^3/\Intr_2\times\Intr_2$ singularities. The aim of this section is to obtain results that hold for all choices of triangulation simultaneously: such results can be uncovered by superimposing the conditions for all the different choices of triangulation. It should be emphasised that we do not wish to imply that it is necessary that such results apply in all triangulations from the supergravity perspective. But surprisingly, superimposing consistency conditions leads to a huge reduction of the complexity of these equations. However, if all consistency conditions are satisfied in any triangulation, then arbitrary flop--transitions are admissible which opens up the possibility to study the resulting transitions in the massless spectra.

\subsection{Flux Quantisation} 

Even though the flux quantisation conditions might seem to be dependent on the choice of the triangulations at the local singularities, in fact, they are all equivalent to 
\equ{ \label{eq:TriangulationFluxQuantisation}
2\, \cV_{i,\gm\gn} \cong 0~, 
\quad 
\sum_\gr \cV_{i,\gr\gn} \cong 0~, 
\quad 
\sum_\gr \cV_{i,\gm\gr} \cong 0~, 
\quad 
\cV_{1,\beta\gamma}+\cV_{2,\alpha\gamma}+\cV_{3,\alpha\beta} \cong 0
}
independently of the local triangulations chosen. To see this, notice first of all that the first three relations derived from curves that exist for any triangulation, see Table~\ref{tb:FluxQuantisation}. Now, if triangulation $E_1$ has been chosen at the resolution of $f_{\ga\gb\gg}$, one has to impose the condition associated to curve $D_{1,\ga} E_{1,\gb\gg}$, if triangulation $E_2$ the  condition associated to curve $D_{2,\gb} E_{2,\ga,\gg}$ and if triangulation $E_3$ the condition associated to curve $D_{3,\gg} E_{3,\ga\gb}$, respectively, while if trangulation $S$ is used all the resulting three conditions have to be superimposed. However, all three of them are equivalent to the last condition in~\eqref{eq:TriangulationFluxQuantisation} using the first condition in this line which basically says that the signs of the bundle vectors in the flux quantisation conditions are irrelevant modulo 2. In other words, if the flux quantisation is satisfied for a single triangulation choice at all the 64 resolved $\Cplx^3/\Intr_2\times\Intr_2$ singularities, the fluxes are properly quantised for any triangulation choice.

\subsection{Reduction of Bianchi Identities}

To determine the set of equations which guarantee that for any choice of triangulation of the 64 $\Cplx^3/\Intr_2\times \Intr_2$ resolutions, the Bianchi identities are solved, we can treat the triangulation dependent functions,  
$\Delta^1_{\alpha\beta\gamma}$, $\Delta^2_{\alpha\beta\gamma}$ and $\Delta^3_{\alpha\beta\gamma}$, as arbitrary functions. Hence, to solve the Bianchi identities for all choices, the coefficients in front of these functions need to cancel among themselves as well as the remaining contributions which do not multiply any of them. This leads to four set of equations for each set of sixteen Bianchi identities on each of the exceptional cycles. 
For the sixteen Bianchi identities on $E_{1,\beta\gamma}$ they read:
\begin{equation}
\begin{array}{c}
\sum\limits_\alpha \Big[ 
\cV_{2,\alpha\gamma}^2 + \cV_{3,\alpha\beta}^2 - \cV_{1,\beta\gamma}^2  
+ 2\, \cV_{1,\beta\gamma} \cdot \cV_{2,\alpha\gamma} + 2\, \cV_{1,\beta\gamma} \cdot \cV_{3,\alpha\beta} 
-2\, \cV_{2,\alpha\gamma} \cdot \cV_{3,\alpha\beta} 
\Big] = -8~,  
\\[3ex] 
\cV_{1,\beta\gamma}^2  + \cV_{2,\alpha\gamma}^2 + \cV_{3,\alpha\beta}^2 
- 2\, \cV_{2,\alpha\gamma} \cdot \cV_{3,\alpha\beta} = 4~, 
\quad 
\cV_{1,\beta\gamma} \cdot \cV_{2,\alpha\gamma} =  \cV_{1,\beta\gamma} \cdot \cV_{3,\alpha\beta} 
= \cV_{2,\alpha\gamma} \cdot \cV_{3,\alpha\beta}~. 
\end{array}
\end{equation}
For the sixteen Bianchi identities on $E_{2,\alpha\gamma}$ they read:
\begin{equation}
\begin{array}{c}
\sum\limits_\beta \Big[ 
\cV_{1,\beta\gamma}^2 + \cV_{3,\alpha\beta}^2 - \cV_{2,\alpha\gamma}^2  
+ 2\, \cV_{2,\alpha\gamma} \cdot \cV_{1,\beta\gamma} + 2\, \cV_{2,\alpha\gamma} \cdot \cV_{3,\alpha\beta} 
-2\, \cV_{1,\beta\gamma} \cdot \cV_{3,\alpha\beta} 
\Big] = -8~,  
\\[3ex] 
\cV_{1,\beta\gamma}^2  + \cV_{2,\alpha\gamma}^2 + \cV_{3,\alpha\beta}^2 
- 2\, \cV_{1,\beta\gamma} \cdot \cV_{3,\alpha\beta} = 4~, 
\quad 
\cV_{1,\beta\gamma} \cdot \cV_{2,\alpha\gamma} =  \cV_{1,\beta\gamma} \cdot \cV_{3,\alpha\beta} 
= \cV_{2,\alpha\gamma} \cdot \cV_{3,\alpha\beta}~. 
\end{array}
\end{equation}
And, finally, for the sixteen Bianchi identities on $E_{3,\alpha\beta}$ they read:
\begin{equation}
\begin{array}{c}
\sum\limits_\gamma \Big[ 
\cV_{1,\beta\gamma}^2 + \cV_{2,\alpha\gamma}^2 - \cV_{3,\alpha\beta}^2  
+ 2\, \cV_{3,\alpha\beta} \cdot \cV_{1,\beta\gamma} + 2\, \cV_{3,\alpha\beta} \cdot \cV_{2,\alpha\gamma} 
-2\, \cV_{1,\beta\gamma} \cdot \cV_{2,\alpha\gamma} 
\Big] = -8~,  
\\[3ex] 
\cV_{1,\beta\gamma}^2  + \cV_{2,\alpha\gamma}^2 + \cV_{3,\alpha\beta}^2 
- 2\, \cV_{1,\beta\gamma} \cdot \cV_{2,\alpha\gamma} = 4~, 
\quad 
\cV_{1,\beta\gamma} \cdot \cV_{2,\alpha\gamma} =  \cV_{1,\beta\gamma} \cdot \cV_{3,\alpha\beta} 
= \cV_{2,\alpha\gamma} \cdot \cV_{3,\alpha\beta}~. 
\end{array}
\end{equation}
Note that every time the top equations have a sum over one of the fixed point labels, while the lower three do not. Fortunately, many of these equations are redundant. The lower three relations for all three exceptional divisors have the same content: for any choice of $(\alpha,\beta,\gamma)$ the three inner products are constraint to: 
\begin{equation}  \label{eq:TriangulationBIsI}
\cV_{1,\beta\gamma} \cdot \cV_{2,\alpha\gamma} =  \cV_{1,\beta\gamma} \cdot \cV_{3,\alpha\beta} 
= \cV_{2,\alpha\gamma} \cdot \cV_{3,\alpha\beta} = 
 \tfrac12\big( \cV_{1,\beta\gamma}^2  +\cV_{2,\alpha\gamma}^2 +\cV_{3,\alpha\beta}^2\big) -2~.
\end{equation}
Inserting these in the top equations with the sums results in $3\cdot 16$ equations 
\begin{equation} 
\sum\limits_\alpha \Big[ \cV_{2,\alpha\gamma}^2 + \cV_{3,\alpha\beta}^2 \Big] = 12~, 
\quad 
\sum\limits_\beta \Big[ \cV_{1,\beta\gamma}^2 + \cV_{3,\alpha\beta}^2 \Big] = 12~, 
\quad 
\sum\limits_\alpha \Big[ \cV_{1,\beta\gamma}^2 + \cV_{2,\alpha\gamma}^2  \Big] = 12~.
\end{equation} 
If these equations are satisfied, then also the three Bianchi identities on the inherited divisors are automatically satisfied. Indeed, if one sums each of these sets of equations over the other two labels and then add two and subtract a third, the inherited Bianchi identities are recovered. But in fact these equations can be reduced even further by a similar procedure: Sum over one of the two free labels in these equations. One of the two terms is then precisely of the form of one of the inherited Bianchi identities equal to 24. Inserting that and rewriting leads to three sets of $2\cdot 4=8$ (hence 24 in total) even simpler equations:
\begin{equation} \label{eq:TriangulationBIsII}
\sum\limits_\beta \cV_{1,\beta\gamma}^2 = \sum\limits_\gamma \cV_{1,\beta\gamma}^2 = 6~,
\quad  
\sum\limits_\alpha \cV_{2,\alpha\gamma}^2 = \sum\limits_\gamma \cV_{2,\alpha\gamma}^2 = 6~, 
\quad  
\sum\limits_\alpha \cV_{3,\alpha\beta}^2 = \sum\limits_\beta \cV_{3,\alpha\beta}^2 =  6~.  
\end{equation} 
Hence, if the equations \eqref{eq:TriangulationBIsI} and  \eqref{eq:TriangulationBIsII} are simultaneously satisfied, then a solution is obtained of the 51 Bianchi identities that holds in any triangulation. In fact, in each of the three sets of 8 equations there is one linear dependence, since summing over the free indices in both (four) equations in each set, leads to the same equation.


\subsection{Blowup Modes Without Oscillator Excitations}

Assuming that all line bundle vectors can be associated to twisted states without oscillators, then they all square to: 
\begin{equation} \label{eq:BundleVectorsTwistedStates} 
\cV_{a,\mu\nu}^2 = \tfrac 32
\qquad\Rightarrow\qquad
\cV_{a,\mu\nu} \cdot \cV_{b,\rho\sigma} = \tfrac 14~; 
\end{equation} 
the equation after the implication sign follows upon using~\eqref{eq:TriangulationBIsI}, with $\nu=\sigma$ for $a=1$ and $b=2$,  $\mu=\sigma$ for $a=1$ and $b=3$, $\mu=\rho$ for $a=2$ and $b=3$, respectively. Since this solves all equations \eqref{eq:TriangulationBIsI} and \eqref{eq:TriangulationBIsII},  such choices solve all Bianchi identities in any triangulation simultaneously.

\subsection{Consequences of Triangulation Independence for the Multiplicity Operator}
\label{sc:ConsequenceTriangulationMultiplicities}

Contrary to the fundamental consistency conditions, the multiplicity operator does not simplify in any particular way, when line bundle resolutions models are considered that are admissible in any choice of triangulation of the 64 $\Intr_2\times\Intr_2$ resolved singularities. However, it can be brought in a specific form. Since the $S$--triangulation plays a special role in flop--transitions as any flop involves the $S$--triangulation, the $S$--triangulation can be taken to be the reference triangulation at all the 64 $\Intr_2\times\Intr_2$ resolved singularities. Using this the total multiplicity operator $\mathsf{N}$ can be written as 
\equ{ \label{eq:TriangulationDependentMultiplicitiesReference} 
\mathsf{N} = 
\mathsf{N}^{S} + 
\sum_{\ga,\gb,\gg} \Big[ 
\gd^{E_1}_{\ga\gb\gg}\, \gD \mathsf{N}^{1}_{\ga\gb\gg} + 
\gd^{E_2}_{\ga\gb\gg}\, \gD \mathsf{N}^{2}_{\ga\gb\gg} + 
\gd^{E_3}_{\ga\gb\gg}\, \gD \mathsf{N}^{3}_{\ga\gb\gg} 
\Big]~, 
\qquad 
\mathsf{N}^S = \sum_{\ga,\gb,\gg} \mathsf{N}^S_{\ga\gb\gg}
}
and $\gD \mathsf{N}^{i}_{\ga\gb\gg}$ defined in~\eqref{eq:DifferenceMultiplicities}. Both $\mathsf{N}^S$ and $\gD \mathsf{N}^{i}_{\ga\gb\gg}$ are always integer: $\mathsf{N}^S$ is the multiplicity operator when at all 64 resolved singularities triangulation $S$ is taken, hence it has to be integral on all chiral states in the spectrum; for the triangulation difference multiplicities $\gD \mathsf{N}^{i}_{\ga\gb\gg}$ it was already established in Subsection~\ref{sc:FlopTransitionSpectraJumps} that they are always integral. This means that in the most general case one can define $3\cdot 4^3+1=193$ multiplicity operators ($\gD\mathsf{N}^{i}_{\ga\gb\gg}$ for $i=1,2,3$, $\ga,\gb,\gg = 1,2,3,4$ and $\mathsf{N}^S$) that all have to be integral on any $E_8\times E_8$ root.

\section{Models Without Wilson Lines}
\label{sc:NoWilsonLines} 

This section is devoted to a number of simple line bundle models to illustrate the main ideas about dealing with the triangulation dependence. The focus is on demonstrating that the approach to parameterise all triangulations in the way discussed in the preceding sections always lead to sensible, {\em e.g.}\ integral spectra for any triangulation chosen provided that the consistency conditions have been solved for all triangulations simultaneously. However, these models should not be considered as fully realistic models. In particular,  the consequences of the DUY equations will be mostly ignored.

\subsection{$\boldsymbol{T^6/\Intr_2\times\Intr_2}$ Orbifold Models}

From the orbifold perspective these are models without Wilson lines, this means that such orbifold models are characterised by just two gauge shifts $V_a$. They satisfy 
\equ{
V_a \cdot V_b - v_a \cdot v_b  \equiv 0~, 
}
for $a,b = 1,2$. Here $v_a$ denote the two independent four--component geometrical $\Intr_2$ orbifold twists and $V_a$ the sixteen--component shift embedding on the gauge lattice taken to be either the weight lattice of $E_8\times E_8$ or $Spin(32)/\Intr_2$. Furthermore, it is often convenient to introduce $v_3 \cong v_1+v_2$ and $V_3 \cong V_1+V_2$.

\begin{table} 
\[
\arry{|c||l|l|l|}{
\hline 
\text{Model} & \multicolumn{3}{c|}{\text{Twists / Gauge Shift Embeddings}}
\\ \hline\hline 
I & v_1 = (0,0, \tfrac 12, -\tfrac 12)~,
&
v_2 = (0,-\tfrac 12, 0,\tfrac 12)~, 
&
v_3 =  (0,\tfrac 12, -\tfrac 12,0)~. 
\\ \hline 
I.a & 
V_1 = (0, \tfrac 12, -\tfrac 12, 0^5)(0^8) 
&
V_2 = (-\tfrac 12, 0,\tfrac 12, 0^5)(0^8)
&
V_3 = (\tfrac 12,-\tfrac 12,0, 0^5)(0^8)
\\
I.b & 
V_1 = (0, \tfrac 12, -\tfrac 12, 0^5)(1,0^7) 
&
V_2 = (-\tfrac 12, 0,\tfrac 12, 0^5)(-1,0^7)
&
V_3 = (\tfrac 12, -\tfrac 12,0, 0^5)(0,0^7)
\\ \hline\hline 
II & 
v_1 = (0,0, \tfrac 12, \tfrac 12)~,
&
v_2 = (0,\tfrac 12, 0,\tfrac 12)~, 
&
 v_3 = (0,-\tfrac 12, -\tfrac 12,1)~, 
\\ \hline 
II.a & 
V_1 = (0, \tfrac 12, \tfrac 12, 0^5)(0^8) 
&
V_2 = (\tfrac 12, 0,\tfrac 12, 0^5)(0^8)
&
V_3 = (-\tfrac 12,-\tfrac 12,1, 0^5)(0^8)
\\
II.b & 
V_1 = (0, \tfrac 12, \tfrac 12, 0^5)(1,0^7) 
&
V_2 = (\tfrac 12, 0, \tfrac 12, 0^5)(-1,0^7)
&
V_3 = (-\tfrac 12, -\tfrac 12, 1, 0^5)(0,0^7)
\\ \hline 
}
\]
\caption{  \label{eq:Z2Z2OrbiGaugeEmbs}
This table gives two different choices of the orbifold twist vectors $v_a$ and two associated 
inequivalent gauge shift embeddings $V_a$ for each choice. 
}
\end{table}

The geometrical twists $v_1$ and $v_2$ are conventionally chosen such as to preserve $\cN=1$ target space supersymmetry. On the level of the orbifold theory there are various equivalent choices for them. The most commonly used choice $I.$\ is 
\begin{subequations} \label{eq:Z2Z2OrbiTwistsGeomI}
\begin{equation} 
v_1 = (0,0, \tfrac 12, -\tfrac 12)~,
\qquad 
v_2 = (0,-\tfrac 12, 0,\tfrac 12)~, 
\qquad
 v_3 = v_1+v_2 = (0,-\tfrac 12, \tfrac 12,0) \cong  (0,\tfrac 12, -\tfrac 12,0)~. 
\end{equation}
The first entry of these vectors corresponds to the four--dimensional Minkowski space in light--cone gauge; the other three components to the twist actions on the three two--torus that make up the $T^6$. The final expression for $v_3$ obtained by adding a lattice vector $(0,1,-1,0)$ making a permutation symmetry between the entries of $v_1, v_2$ and $v_3$ manifest. Note that with this form of $v_3$ 
\equ{
v_a \cdot v_b = -\tfrac 14 + \tfrac 34\, \gd_{ab}~. 
}
\end{subequations} 
A second choice $II.$\ is given by 
\begin{subequations}  \label{eq:Z2Z2OrbiTwistsGeomII}
\begin{equation} 
v_1 = (0,0, \tfrac 12, \tfrac 12)~,
\qquad 
v_2 = (0,\tfrac 12, 0,\tfrac 12)~, 
\qquad
 v_3 = v_1+v_2 = (0,\tfrac 12, \tfrac 12,1) \cong (0,-\tfrac 12, -\tfrac 12,1)~, 
\end{equation}
where the latter form of $v_3$ in this case is obtained by adding $(0,-1,-1,0)$. With this form of $v_3$ the vectors $v_a$ satisfy
\equ{
v_a \cdot v_b = \tfrac 14\big(1 +  \gd_{ab}\big) + \gd_{a3}\,\gd_{b3}~. 
}
\end{subequations} 
On the level of the orbifold theory both choices are equivalent. For both these forms there are two inequivalent choices for the gauge embedding, denoted by $a$ and $b$. This leads to four possible gauge shift embeddings referred to as $I.a$, $I.b$, $II.a$ and $II.b$ in Table~\ref{eq:Z2Z2OrbiGaugeEmbs}. On the level of the orbifold only the choices $a$ or $b$ lead to physically different models; as is shown below on the level of the resolution the choice of twist $I.$ or $II.$ is of significance as only one of the two choices can be associated to a line bundle model.

\subsection{Models with Three Independent Line Bundles}

First some general facts  about associated blowup models are given. In this section the line bundle vectors are taken to be independent of the labels $\ga,\gb,\gg$, {\em i.e.}: 
\equ{
\cV_{1,\gb\gg} = \cV_1~, 
\quad 
\cV_{2,\ga\gg} = \cV_2~, 
\quad 
\cV_{3,\ga\gb} = \cV_3~. 
}
Consequently the triangulation independent flux quantisation conditions~\eqref{eq:TriangulationFluxQuantisation} reduce to 
\equ{ \label{eq:FluxQuant3Bundles} 
2\, \cV_1 \cong 2\, \cV_2 \cong 2\, \cV_3 \cong \cV_1 + \cV_2 + \cV_3 \cong 0~. 
}
Such bundle vectors, $\cV_1, \cV_2$ and $\cV_3$, can be obtained from the orbifold gauge shift vectors $V_1, V_2$ and $V_3$,  by adding appropriate lattice vectors $L_{1,\beta\gamma}, L_{2,\alpha\gamma}$ and $L_{3,\alpha\beta}$. In this section they are chosen such that the bundle vectors $\cV_1, \cV_2$ and $\cV_3$ can be associated with twisted states without oscillators satisfying the conditions~\eqref{eq:BundleVectorsTwistedStates}.

The number  $N_T$ of times, that triangulation $T=E_1$, $E_2$, $E_3$, $S$ has been chosen at the 64 resolved $\Cplx^3/\Intr_2\times\Intr_2$ singularities, can be determined by summing the functions $\gd^T_{\ga\gb\gg}$ over all of them, {\em e.g.}: 
\equ{
N_{T} = \sum_{\ga,\gb,\gg} \gd^T_{\ga\gb\gg}~, 
}
hence, in particularly, for $i=1,2,3$: 
\equ{ \label{eq:SumTriangNums}
\sum_{\ga,\gb,\gg} \Big( 1 - \gD^i_{\ga\gb\gg} \Big) = 
2\, N_{E_i}  + N_{S}~,
\qquad
N_E + N_{S} = 64~, 
\qquad 
N_E = N_{E_1} + N_{E_2} + N_{E_3}~. 
}
Then, if also the Cartan operators are abbreviated as   
\equ{
\mathsf{H}_{1} = \mathsf{H}_{1,\gb\gg}~, 
\qquad 
\mathsf{H}_{2}  =  \mathsf{H}_{2,\ga\gg} ~, 
\qquad 
 \mathsf{H}_{3} =  \mathsf{H}_{3,\ga\gb}~, 
}
the multiplicity operator~\eqref{eq:TriangulationDependentMultiplicities}  simplifies to 
\equ{ \label{eq:TotalMultiplicities}
\mathsf{N} =
N_{E_1}\,\mathsf{N}^{1} + 
N_{E_2}\, \mathsf{N}^{2} + 
N_{E_3}\, \mathsf{N}^{3} + 
N_{S}\, \mathsf{N}^{S}~, 
} 
expressed in terms of four multiplicity operators for each of the four triangulations
\begin{subequations} \label{eq:TriangulationMultiplicities}
\equa{
 \mathsf{N}^1 &= 
\tfrac 14\, \mathsf{H}_{1} 
+ \tfrac 1{12}\, \mathsf{H}_{2} \big( 4 \mathsf{H}_{2}^2-1\big) 
+ \tfrac 1{12}\, \mathsf{H}_{3}\big( 4 \mathsf{H}_{3}^2-1\big) 
- \mathsf{H}_{1} \big( \mathsf{H}_{2}^2 + \mathsf{H}_{3}^2 \big)~, 
\\[1ex] 
 \mathsf{N}^2 &= 
\tfrac 14\, \mathsf{H}_{2} 
+ \tfrac 1{12}\, \mathsf{H}_{1} \big( 4 \mathsf{H}_{1}^2-1\big) 
+ \tfrac 1{12}\, \mathsf{H}_{3} \big( 4 \mathsf{H}_{3}^2-1\big) 
- \mathsf{H}_{2} \big( \mathsf{H}_{1}^2 + \mathsf{H}_{3}^2 \big)~, 
\\[1ex] 
 \mathsf{N}^3 &= 
\tfrac 14\, \mathsf{H}_{3} 
+\tfrac 1{12}\, \mathsf{H}_{1} \big( 4 \mathsf{H}_{1}^2-1\big) 
+ \tfrac 1{12}\, \mathsf{H}_{2} \big( 4 \mathsf{H}_{2}^2-1\big) 
- \mathsf{H}_{3} \big( \mathsf{H}_{1}^2 + \mathsf{H}_{2}^2 \big)~, 
\\[1ex] 
 \mathsf{N}^S &= \non 
\tfrac 1{12}\, \mathsf{H}_{1}\big( 2\mathsf{H}_{1}^2 +1) 
+\tfrac 1{12}\, \mathsf{H}_{2}\big( 2\mathsf{H}_{2}^2 +1) 
+\tfrac 1{12}\, \mathsf{H}_{3}\big( 2\mathsf{H}_{3}^2 +1) 
+\mathsf{H}_{1} \mathsf{H}_{2} \mathsf{H}_{3} 
\\[1ex] 
&\phantom{=} 
-\tfrac 12\, \mathsf{H}_{1}\big(\mathsf{H}_{2}^2 + \mathsf{H}_{3}^2\big)
-\tfrac 12\, \mathsf{H}_{2
} \big( \mathsf{H}_{1}^2 + \mathsf{H}_{3}^2\big)
 -\tfrac 12\, \mathsf{H}_{3} \big( \mathsf{H}_{1}^2+\mathsf{H}_{2}^2\big)~. 
}
\end{subequations} 
Since $N_{E_1}$, $N_{E_2}$, $N_{E_3}$ and $N_{S}$ are arbitrary non--negative integers subject to~\eqref{eq:SumTriangNums}, it follows that if we substitute one of them away in~\eqref{eq:TotalMultiplicities}, the resulting expression has to be integral on all $E_8\times E_8$ weights for any choice of the remaining numbers. In particular, taking the triangulation $S$ again as reference, {\em i.e.}\ solving $N_S$ from~\eqref{eq:SumTriangNums} and substituting this in~\eqref{eq:TotalMultiplicities}, gives 
\equ{ \label{eq:TotalMultiplicitiesII}
\mathsf{N} =
N_{E_1}\,\gD\mathsf{N}^{1} + 
N_{E_2}\, \gD\mathsf{N}^{2} + 
N_{E_3}\, \gD\mathsf{N}^{3} + 
64\, \mathsf{N}^{S}~, 
\qquad 
\gD\mathsf{N}^{i} = \mathsf{N}^{i}-\mathsf{N}^{S}~,
} 
for $i=1,2,3$. In line with the general observation in Section~\ref{sc:ConsequenceTriangulationMultiplicities}, this expression should always be integral. Hence, in particular, the operators $\gD\mathsf{N}^i$ have to be integral on any state.

\subsection{$\boldsymbol{SO(10)\times SO(12)}$ Line Bundle Models}
\label{sc:SO1012models}

Starting from the orbifold gauge embeddings $II.b$ of the classification in Table~\ref{eq:Z2Z2OrbiGaugeEmbs} a set of three line bundle vectors can be obtained 
\begin{subequations} \label{eq:SO1012Bundles}
\begin{eqnarray}
\cV_1 = V_1 + L_{1}= (0,\tfrac 12,\tfrac 12, 0^5)(1,0,0^6)~,
& \qquad &
L_{1} = (0^8)(0^8)
\\[1ex]  
\cV_2 = V_2 + L_{2} = (\tfrac 12,0,\tfrac 12,0^5)(0,1,0^6)~,
& \qquad &
L_{2} = (0^8)(-1,1,0^6)~,
\\[1ex]  
\cV_3 =  V_3 + L_{3} = (-\tfrac 12,-\tfrac 12,1,0^5)(0,0,0^6)~,
& \qquad &
L_{3} = (0^8)(0^8)~. 
\end{eqnarray}
\end{subequations}
The bundle vectors satisfy the flux quantisation~\eqref{eq:FluxQuant3Bundles} for arbitrary triangulations. Note that, these bundle vectors cannot be obtained from orbifold model $I.b$\,. Thus equivalent choices on the orbifold level might lead to inequivalent choices from the smooth resolved perspective. The unbroken non--Abelian gauge group is $SO(10)\times SO(12)$.

\begin{table}
\[
\arry{|c||ccc|cccc|ccc|c|}{
\hline 
\text{weight} & \mathsf{H}_1 & \mathsf{H}_2 & \mathsf{H}_3 & 
\mathsf{N}^1 & \mathsf{N}^2 &\mathsf{N}^3 &\mathsf{N}^{S} &
\gD\mathsf{N}^{1} & \gD\mathsf{N}^{2} & \gD\mathsf{N}^{3} & 
\text{repr.}
\\ \hline\hline 
(1,0,0, \undr{\pm1,0^4})(0^8) & 0 & \tfrac 12 & -\tfrac 12  
&0 & 0 & 0 & 0 & 0 & 0 & 0 
& (\rep{10})(\rep{1})  
\\ 
(0,1,0, \undr{\pm1,0^4})(0^8) & \tfrac 12 & 0 & -\tfrac 12  
& 0 & 0 & 0 & 0 & 0 & 0 & 0
& (\rep{10})(\rep{1})  
\\ 
(0,0,1, \undr{\pm1,0^4})(0^8) & \tfrac 12 & \tfrac 12 & 1  
& -\tfrac14 & -\tfrac14 & -\tfrac14 & -\tfrac14 & 0 & 0 & 0 
& (\rep{10})(\rep{1})  
\\ \hline
(\tfrac 12,\tfrac 12,\tfrac12,\undr{-\tfrac12^e,\tfrac12^{5-e}})(0^8) 
& \tfrac 12 & \tfrac 12 & 0  
& 0 & 0 & 0 & 0 & 0 & 0 & 0
& (\rep{16})(\rep{1})  
\\
(-\tfrac 12,\tfrac 12,\tfrac12,\undr{-\tfrac12^o,\tfrac12^{5-o}})(0^8) 
& \tfrac 12 & 0 & \tfrac 12 
& 0 & 0 & 0 & 0 & 0 & 0 & 0
& (\crep{16})(\rep{1})  
\\
(\tfrac 12,-\tfrac 12,\tfrac12,\undr{-\tfrac12^o,\tfrac12^{5-o}})(0^8) 
& 0 & \tfrac 12 & \tfrac 12  
& 0 & 0 & 0 & 0 & 0 & 0 & 0
& (\crep{16})(\rep{1})  
\\
(\tfrac 12,\tfrac 12,-\tfrac12,\undr{-\tfrac12^o,\tfrac12^{5-o}})(0^8) 
& 0 & 0 & -1  
& -\tfrac14 & -\tfrac14 & -\tfrac14 & -\tfrac14 & 0 & 0 & 0
& (\crep{16})(\rep{1})  
\\ \hline 
(1,1,0,0^5)(0^8) & \tfrac 12 & \tfrac 12 & -1
& -\tfrac34 & -\tfrac34 & \tfrac14 & -\tfrac34 & 0 & 0 & 1
& (\rep{1})(\rep{1})  
\\ 
(1,0,1,0^5)(0^8) & \tfrac 12 & 1 & \tfrac 12 
& -\tfrac14 & -\tfrac14 & -\tfrac14 & -\tfrac14 & 0 & 0 & 0
& (\rep{1})(\rep{1})  
\\ 
(0,1,1,0^5)(0^8) & 1 & \tfrac 12 & \tfrac 12  
& -\tfrac14 & -\tfrac14 & -\tfrac14 & -\tfrac14 & 0 & 0 & 0
& (\rep{1})(\rep{1})  
\\ \hline 
(1,-1,0,0^5)(0^8) & -\tfrac 12 & \tfrac 12 & 0  
& 0 & 0 & 0 & 0 & 0 & 0 & 0
& (\rep{1})(\rep{1})  
\\ 
(1,0,-1,0^5)(0^8) & -\tfrac 12 & 0 & - \tfrac 32
& 0 & -1 & 0 & 0 & 0 & -1 & 0
& (\rep{1})(\rep{1})  
\\ 
(0,1,-1,0^5)(0^8) & 0 & -\tfrac 12 & -\tfrac 32  
& -1 & 0 & 0 & 0 & -1 & 0 & 0
& (\rep{1})(\rep{1})  
\\ \hline\hline 
(0^8)(1,0,\undr{\pm 1,0^5}) & 1 & 0 & 0
& \tfrac14 & \tfrac14 & \tfrac14 & \tfrac14 & 0 & 0 & 0
& (\rep{1})(\rep{12})
\\ 
(0^8)(0,1,\undr{\pm 1,0^5}) & 0 & 1 & 0 
& \tfrac14 & \tfrac14 & \tfrac14 & \tfrac14 & 0 & 0 & 0
& (\rep{1})(\rep{12})
\\ \hline 
(0^8)(\tfrac 12,\tfrac12,\undr{-\tfrac12^e,\tfrac12^{6-e}}) & \tfrac 12 & \tfrac 12 & 0 
& 0 & 0 & 0 & 0 & 0 & 0 & 0
& (\rep{1})(\rep{32})
\\ 
(0^8)(\tfrac 12,-\tfrac12,\undr{-\tfrac12^o,\tfrac12^{6-o}}) & \tfrac 12 & -\tfrac 12 & 0 
& 0 & 0 & 0 & 0 & 0 & 0 & 0
& (\rep{1})(\crep{32})
\\ \hline 
(0^8)(1,1,0^6) & 1 & 1 & 0 
& -\tfrac12 & -\tfrac12 & \tfrac12 & -\tfrac12 & 0 & 0 & 1
& (\rep{1})(\rep{1})
\\ 
(0^8)(1,-1,0^6) & 1 & -1 & 0  
& -1 & 1 & 0 & 0 & -1 & 1 & 0
& (\rep{1})(\rep{1})
\\ \hline 
}
\]
\caption{ \label{tb:SO1012ChargedStates} 
The line bundle charges $\mathsf{H}_i$, the triangulation multiplicities $\mathsf{N}^i$, $\mathsf{N}^S$ and the difference multiplicities $\gD\mathsf{N}^i$ are given for all the $E_8\times E_8$ roots charged under the line bundle background defined by~\eqref{eq:SO1012Bundles}. The underline of some of the entries in these roots denote all possible permutations of them. Notice, that these difference multiplicities, that measure jumps in the spectrum when going through a flop--transition, are always integral. 
}
\end{table}

The line bundle charges $\mathsf{H}_i$, the triangulation multiplicities $\mathsf{N}^i$, $\mathsf{N}^S$ and the  triangulation difference multiplicities $\gD\mathsf{N}^i$ of all the $E_8\times E_8$ roots are given in Table~\ref{tb:SO1012ChargedStates}. The triangulation multiplicities, and $\mathsf{N}^S$ in particular, are not integrally quantised. This might seem problematic, but it is not: triangulation $S$ can be taken to be the reference triangulation at all 64 resolved singularities. Hence, if triangulation $S$ is chosen at all resolved singularities, the spectrum is 64 times the triangulation multiplicity $\mathsf{N}^S$ and all states come in multiples of 16. Now, if at a certain resolved singularities one of the exceptional triangulations is used then the spectrum always changes by an integral amount as the triangulation difference multiplicities $\gD\mathsf{N}^i$ are integral, see Table~\ref{tb:SO1012ChargedStates}. Indeed, using this table the full spectrum in any triangulation can be determined to be: 
\equ{  \non 
16\big\{ (\rep{10})(\rep{1})_{0,0,\sm 2;0} + 
 (\rep{16})(\rep{1})_{\sm 1,\sm 1,1;0} + 
 (\rep{1})(\rep{12})_{0;2,0} + 
 (\rep{1})(\rep{12})_{0;0,2} + 
 (\rep{1})(\rep{1})_{\sm 2,0,\sm 2;0} + 
 (\rep{1})(\rep{1})_{0,\sm 2,\sm 2;0} 
\big\}+ 
\\[2ex] 
48\, (\rep{1})(\rep{1})_{\sm 2,\sm 2,0;0} + N_{E_3}\, (\rep{1})(\rep{1})_{2,2,0;0}+
N_{E_2}\, (\rep{1})(\rep{1})_{\sm 2,0,2;0} + 
N_{E_1}\, (\rep{1})(\rep{1})_{0,\sm 2,2;0} + 
\\[2ex] \non 
32\, (\rep{1})(\rep{1})_{0;\sm 2,\sm 2} + N_{E_3}\, (\rep{1})(\rep{1})_{0;2,2} + 
N_{E_1}\, (\rep{1})(\rep{1})_{0;\sm 2,2} + N_{E_2}\, (\rep{1})(\rep{1})_{0;2,\sm 2}~.
}
The five $U(1)$ charges given here are two times the first three weight entries of the observable $E_8$ and the first two of the hidden $E_8$. They can be used to distinguish otherwise vector--like states. When triangulation $S$ is chosen at all fixed points, {\em e.g.} $N_{E_1}=N_{E_2}=N_{E_3}=0$\,, the spectrum does not contain any vector--like pairs. For most other choices vector--like pairs do arise, but they presumably acquire a mass at some stage in the effective field theory description.

\subsection{A ``swampland'' $\boldsymbol{SO(10)\times SO(10)}$ models}
\label{sc:SO1010models}

A seemingly closely related model with three independent bundle vectors is given by
\begin{subequations} \label{eq:SO1010Bundles}
\begin{eqnarray}
\cV_{1,\beta\gamma}= \cV_1 = (0,\tfrac 12,\tfrac 12,0^5)(-1,0,0,0^5)~,
\\[1ex]  
\cV_{2,\alpha\gamma}=\cV_2 = (\tfrac 12,0,\tfrac 12,0^5)(0,-1,0,0^5)~,
\\[1ex]  
\cV_{3,\alpha\beta} = \cV_3 = (\tfrac 12,\tfrac 12,0,0^5)(0,0,-1,0^5)~.
\end{eqnarray}
\end{subequations}
This leads to a gauge group $SO(10)\times SO(10)$. The unbroken roots are given by $(0^3,\undr{\pm 1,\pm 1, 0^3})(0^8)$ and $(0^8)(0^3,\undr{\pm 1,\pm 1, 0^3})$. At first sight this seems to be a valid choice as well, but this model has a number of issues:

First of all, even thought the bundle vectors clearly satisfy the strong conditions~\eqref{eq:BundleVectorsTwistedStates}, this model cannot be obtained as the blowup of any orbifold model. The first two bundle vectors are identical to the model discussed in the previous subsection and can be obtained from orbifold gauge shift vectors given there. But the third one does not differ by a lattice vector from $V_1+V_2$: 
\equ{
\cV_3 - V_1 - V_2 = 
(\tfrac 12,\tfrac 12,0,0^5)(0,0,-1,0^5) -  (\tfrac 12, \tfrac 12, 1, 0^5)(0,0,0,0^5) = 
(0,0,-1,0^5)(0,0,-1,0^5)~. 
}
(If both $-1$--entries would have lain in the same $E_8$, this would be a lattice vector, but they don't.) 

Moreover, this choice of line bundle vectors does not satisfy the final flux quantisation condition in~\eqref{eq:FluxQuant3Bundles}. As a consequence, the spectrum is not integral for a generic choice of triangulation at the 64 $\Cplx^3/\Intr_2\times \Intr_2$ resolutions. This can be inferred from the appearance of multiplicities $\pm 1/16$ and $-5/16$ for the $\Delta \mathsf{N}^1$, $\Delta \mathsf{N}^2$ and $\Delta \mathsf{N}^3$ in  Table~\ref{tb:SO1010ChargedStates} when the states are distinguished by their (implicitly given) $U(1)$ charges. Even if one ignores the $U(1)$ charges, the spectrum combined is not necessarily integral: 
\equ{
16\, (\crep{16})(\rep{1}) + 48\, (\rep{1})(\rep{10}) + 4\, (\rep{1})(\rep{16}) + 36\,(\rep{1})(\crep{16}) 
+ \tfrac 18 N_E \big\{  (\rep{16})(\rep{1}) + (\crep{16})(\rep{1}) + 4\, (\rep{1})(\rep{16}) \big\} 
+ \text{singlets}~. 
}
Note, that if the same triangulation is chosen at all 64 resolved $\Cplx^3/\Intr_2\times\Intr_2$ singularities, the spectrum would be integral. But any single flop--transition would then lead to an inconsistent spectrum. This demonstrates that satisfying the flux quantisation conditions in any triangulation is essential for the difference multiplicities $\gD \mathsf{N}^i$ to be always integral.

\begin{table}
\[
\arry{|c||ccc|cccc|ccc|c|}{
\hline 
\text{weight} & \mathsf{H}_1 & \mathsf{H}_2 & \mathsf{H}_3 & 
\mathsf{N}^1 & \mathsf{N}^2 &\mathsf{N}^3 &\mathsf{N}^{S} &
\gD\mathsf{N}^{1} & \gD\mathsf{N}^{2} & \gD\mathsf{N}^{3} & 
\text{repr.}
\\ \hline\hline 
(-\tfrac 12,-\tfrac 12,-\tfrac12,\undr{-\tfrac12^o,\tfrac12^{5-o}})(0^8) 
& \tfrac 12 & \tfrac 12 & 0  
& \tfrac 18 & \tfrac 18  & \tfrac 18  & \tfrac 1{16} & \tfrac 1{16} & \tfrac 1{16} & \tfrac 1{16}
& (\crep{16})(\rep{1})  
\\
(-\tfrac 12,\tfrac 12,\tfrac12,\undr{-\tfrac12^o,\tfrac12^{5-o}})(0^8) 
& \tfrac 12 & 0 & 0 
& \tfrac 18 & 0 & 0 & \tfrac 1{16} & \tfrac 1{16} & -\tfrac 1{16} & -\tfrac 1{16}
& (\crep{16})(\rep{1})  
\\
(\tfrac 12,-\tfrac 12,\tfrac12,\undr{-\tfrac12^o,\tfrac12^{5-o}})(0^8) 
& 0 & \tfrac 12 & 0  
& 0 & \tfrac 18 & 0 & \tfrac 1{16} & -\tfrac 1{16} & \tfrac 1{16} & -\tfrac 1{16}
& (\crep{16})(\rep{1})  
\\
(\tfrac 12,\tfrac 12,-\tfrac12,\undr{-\tfrac12^o,\tfrac12^{5-o}})(0^8) 
& 0 & 0 & \tfrac12  
& 0 & 0 & \tfrac 18 & \tfrac 1{16} & -\tfrac 1{16} & -\tfrac 1{16} & \tfrac 1{16}
& (\crep{16})(\rep{1})  
\\ \hline 
(-1,-1,0,0^5)(0^8) & -\tfrac 12 & -\tfrac 12 & -1
& \tfrac14 & \tfrac14 & \tfrac14 & \tfrac14 & 0 & 0 & 0
& (\rep{1})(\rep{1})  
\\ 
(-1,0,-1,0^5)(0^8) & -\tfrac 12 & -1 & -\tfrac 12 
& \tfrac14 & \tfrac14 & \tfrac14 & \tfrac14 & 0 & 0 & 0
& (\rep{1})(\rep{1})  
\\ 
(0,-1,-1,0^5)(0^8) & -1 & -\tfrac 12 & -\tfrac 12  
& \tfrac14 & \tfrac14 & \tfrac14 & \tfrac14 & 0 & 0 & 0
& (\rep{1})(\rep{1})  
\\ \hline\hline 
(0^8)(-1,0,0,\undr{\pm 1,0^4}) & 1 & 0 & 0
& \tfrac14 & \tfrac14 & \tfrac14 & \tfrac14 & 0 & 0 & 0
& (\rep{1})(\rep{10})
\\ 
(0^8)(0,-1,0,\undr{\pm 1,0^4}) & 0 & 1 & 0 
& \tfrac14 & \tfrac14 & \tfrac14 & \tfrac14 & 0 & 0 & 0
& (\rep{1})(\rep{10})
\\ 
(0^8)(0,0,-1,\undr{\pm 1,0^4}) & 0 & 0 & 1
& \tfrac14 & \tfrac14 & \tfrac14 & \tfrac14 & 0 & 0 & 0
& (\rep{1})(\rep{10})
\\ \hline 
(0^8)(\tfrac 12,\tfrac12,\tfrac12, \undr{-\tfrac12^e,\tfrac12^{5-e}}) & -\tfrac 12 & -\tfrac 12 & -\tfrac 12 
& \tfrac 18 & \tfrac 18 & \tfrac 18 & \tfrac 1{16} & \tfrac 1{16} & \tfrac 1{16} & \tfrac 1{16}
& (\rep{1})(\rep{16})
\\ \hline 
(0^8)(-\tfrac 12,\tfrac12,\tfrac 12,\undr{-\tfrac12^o,\tfrac12^{5-o}}) & \tfrac 12 & -\tfrac 12 & -\tfrac 12 
& -\tfrac 18 & \tfrac 18 & \tfrac 18 & \tfrac 3{16} & -\tfrac 5{16} & -\tfrac 1{16} & -\tfrac 1{16}
& (\rep{1})(\crep{16})
\\
0^8)(\tfrac 12,-\tfrac12,\tfrac 12,\undr{-\tfrac12^o,\tfrac12^{5-o}}) & -\tfrac 12 & \tfrac 12 & -\tfrac 12 
& \tfrac 18 & -\tfrac 18 & \tfrac 18 & \tfrac 3{16} & -\tfrac 1{16} & -\tfrac 5{16} & -\tfrac 1{16}
& (\rep{1})(\crep{16})
\\
0^8)(\tfrac 12,\tfrac12,-\tfrac 12,\undr{-\tfrac12^o,\tfrac12^{5-o}}) & \tfrac 12 & -\tfrac 12 & -\tfrac 12 
& \tfrac 18 & \tfrac 18 & -\tfrac 18 & \tfrac 3{16} & -\tfrac 1{16} & -\tfrac 1{16} & -\tfrac 5{16}
& (\rep{1})(\crep{16})
\\ \hline 
(0^8)(0,1,1,0^5) & 0 & -1 & -1 
& -\tfrac12 & \tfrac12 & \tfrac12 & \tfrac12 & -1 & 0 & 0
& (\rep{1})(\rep{1})
\\ 
(0^8)(1,0,1,0^5) & -1 & 0 & -1 
& \tfrac12 & -\tfrac12 & \tfrac12 & \tfrac12 & 0 & -1 & 0
& (\rep{1})(\rep{1})
\\ 
(0^8)(1,1,0,0^5) & -1 & -1 & 0 
& \tfrac12 & \tfrac12 & -\tfrac12 & \tfrac12 & 0 & 0 & -1
& (\rep{1})(\rep{1})
\\ \hline 
(0^8)(1,-1,0,0^5) & -1 & 1 & 0  
& 1 & -1 & 0 & 0 & 1 & -1 & 0
& (\rep{1})(\rep{1})
\\
(0^8)(-1,0,1,0^5) & 1 & 0 & -1  
& -1 & 0 & 1 & 0 & -1 & 0 & 1
& (\rep{1})(\rep{1})
\\
(0^8)(0,1,-1,0^5) & 0 & 1 & -1 
& 0 & -1 & 1 & 0 & 0 & -1 & 1
& (\rep{1})(\rep{1})
\\ \hline 
}
\]
\caption{ \label{tb:SO1010ChargedStates} 
The line bundle charges $\mathsf{H}_i$ and the triangulation multiplicities $\mathsf{N}^i$, $\mathsf{N}^S$  are given for all the $E_8\times E_8$ roots charged under the line bundle background defined by~\eqref{eq:SO1010Bundles}. Note that for this model many of the triangulation difference multiplicities $\gD\mathsf{N}^i$ are non--integral signifying that this model is not fully consistent. 
}
\end{table}

\subsection{Blaszczyk's $\boldsymbol{SU(3)\times SU(2)}$ Line Bundle Models}
\label{sc:SU32models} 

An example with very similar line bundle vectors, but where all their non--trivial entries lie in the observable $E_8$ can be obtained from the orbifold gauge embeddings $I.a$ of Table~\ref{eq:Z2Z2OrbiGaugeEmbs}. (But these bundle vectors cannot be obtained from orbifold model $II.a$\,.) The defining set of three line bundle vectors are given by: 
\begin{subequations} \label{eq:SU32Bundles}
\begin{eqnarray}
\cV_1 = V_1 + L_{1}= (0,\tfrac 12,\tfrac 12,-1,0,0, 0^2)(0^8)~,
& \qquad &
L_{1} = (0,0,1,-1,0,0,0^2)(0^8)
\\[1ex]  
\cV_2 = V_2 + L_{2} = (\tfrac 12,0,\tfrac 12,0,-1,0,0^2)(0^8)~,
& \qquad &
L_{2} = (0,1,0,0,-1,0,0^2)(0^8)~,
\\[1ex]  
\cV_3 =  V_3 + L_{3} = (\tfrac 12,\tfrac 12,0,0,0,-1,0^2)(0^8)~,
& \qquad &
L_{3} = (1,0,0,0,0,-1,0^2)(0^8)~, 
\end{eqnarray}
\end{subequations}
These bundle vectors were considered in Section 4.3 of ref.~\cite{Blaszczyk:2010db} before. In that work the spectra were obtained when at all 64 resolved $\Cplx^3/\Intr_2\times\Intr_2$ singularities one of the four possible triangulations were chosen. However, they satisfy the very restrictive conditions~\eqref{eq:BundleVectorsTwistedStates} that ensures that the Bianchi identities are satisfied and the flux quantisation conditions~\eqref{eq:FluxQuant3Bundles} for all triangulations simultaneously. Hence, this set of bundle vectors do not suffer from the flaws encountered in the section above.

Besides all the hidden $E_8$ roots there are six unbroken $SU(3)$ roots $\pm (0^6,1^2)$ and $\pm (\tfrac 12^6, \pm \tfrac 12^2)(0^8)$ and two unbroken $SU(2)$ roots $\pm(0^6,1,-1)$, consequently the unbroken non--Abelian gauge group is $SU(3)\times SU(2) \times E_8$. The Cartan generators of $SU(3)$ are $h_1=(\tfrac 12^6,-\tfrac 12^2)$ and $h_2 = (0^6,1^2)$ and of $SU(2)$ $h=(0^6,1,-1)$.

The triangulation multiplicities evaluated on all observable $E_8$ roots are given in Table~\ref{tb:SU32ChargedVectorStates}. If the same triangulation is used at all resolved singularities then the spectra given in Table 10 of  ref.~\cite{Blaszczyk:2010db} are reproduced. But with the formalism laid out in this paper an arbitrary triangulation of each of the resolved fixed points can be considered. As the triangulation difference multiplicities $\gD \mathsf{N}^i$ are all integral and the states come in multiples of 16 if triangulation $S$ is used at all resolved singularities, the spectrum is integral for any choice of local triangulations. Indeed, ignoring all $U(1)$ charges, the full charged $SU(3)\times SU(2)$ spectrum from the observable $E_8$ can be compactly summarised as 
\equ{
48\, (\rep{3},\rep{2}) + 96\, (\crep{3},\rep{1}) + (96 + N_E)\,\big\{ (\crep{3},\rep{1}) + (\rep{3},\rep{1})\big\} 
+ (176-2\, N_E)\, (\rep{1},\rep{2}) + (144 + N_E)\, (\rep{1},\rep{1})~. 
}
It can be easily confirmed from this spectrum that $SU(3)$ cubed anomaly cancels for any $N_E$ and the $SU(2)$ Witten anomaly is always absent since the number of $SU(2)$ doublets is always even.


\begin{sidewaystable}
\begin{center} 
\scalebox{0.6675}{\(
\begin{array}[t]{cc} 
\arry{|c||ccc|cccc|ccc|c|}{
\hline 
\text{weight} & \mathsf{H}_1 & \mathsf{H}_2 & \mathsf{H}_3 & 
\mathsf{N}^1 & \mathsf{N}^2 &\mathsf{N}^3 &\mathsf{N}^{S} &
\gD\mathsf{N}^{1} & \gD\mathsf{N}^{2} & \gD\mathsf{N}^{3} & 
\text{repr.}
\\ \hline\hline 
(\tfrac 12,\tfrac 12,\tfrac12,-\tfrac 12,\tfrac 12, \tfrac 12, \undr{-\tfrac12,\tfrac12})  &
1 & 0 & 0 & \tfrac 14 & \tfrac 14 & \tfrac 14 & \tfrac 14 & 0 & 0 & 0
& (\rep{3},\rep{2})  
\\
(0,0,0,-1,0,0, \undr{\pm1,0})  & 
&&&&&&&&& 
& 
\\ \hdashline 
(\tfrac 12,\tfrac 12,\tfrac12,\tfrac 12,-\tfrac 12, \tfrac 12, \undr{-\tfrac12,\tfrac12})  &
0 & 1 & 0 & \tfrac 14 & \tfrac 14 & \tfrac 14 & \tfrac 14 & 0 & 0 & 0
& (\rep{3},\rep{2})  
\\
(0,0,0,0,-1,0, \undr{\pm1,0})  & 
&&&&&&&&& 
&  
\\ \hdashline 
(\tfrac 12,\tfrac 12,\tfrac12,\tfrac 12,\tfrac 12, -\tfrac 12, \undr{-\tfrac12,\tfrac12})  & 
0 & 0 & 1 & \tfrac 14 & \tfrac 14 & \tfrac 14 & \tfrac 14 & 0 & 0 & 0
& (\rep{3},\rep{2})  
\\ 
(0,0,0,0,0,-1, \undr{\pm1,0})  & 
&&&&&&&&& 
&  
\\ \hline 
(-1,-1,0,0,0,0, 0^2)  & 
-\tfrac 12 & -\tfrac 12 & -1 & \tfrac 14 & \tfrac 14 & \tfrac 14 & \tfrac 14 & 0 & 0 & 0
& (\crep{3},\rep{1})  
\\ 
(-\tfrac 12,-\tfrac 12,\tfrac12,\tfrac 12,\tfrac 12, \tfrac 12,\pm\tfrac12^2)  & 
&&&&&&&&& 
& 
\\ \hdashline 
(-1,0,-1,0,0,0, 0^2)  & 
-\tfrac 12 & -1 & -\tfrac 12 & \tfrac 14 & \tfrac 14 & \tfrac 14 & \tfrac 14 & 0 & 0 & 0
& (\crep{3},\rep{1})  
\\ 
(-\tfrac 12,\tfrac 12,-\tfrac12,\tfrac 12,\tfrac 12, \tfrac 12,\pm\tfrac12^2)  &
&&&&&&&&& 
& 
\\ \hdashline 
(0,-1,-1,0,0,0, 0^2)  & 
-1 & -\tfrac 12 & -\tfrac 12 & \tfrac 14 & \tfrac 14 & \tfrac 14 & \tfrac 14 & 0 & 0 & 0
& (\crep{3},\rep{1})  
\\
(\tfrac 12,-\tfrac 12,-\tfrac12,\tfrac 12,\tfrac 12, \tfrac 12,\pm\tfrac12^2)  & 
&&&&&&&&& 
& 
\\ \hline  
(-1,0,0,-1,0,0, 0^2)  & 
1 & -\tfrac 12 & -\tfrac 12 & -\tfrac 14 & \tfrac 34 & \tfrac 34 & \tfrac 34 & -1 & 0 & 0
& (\crep{3},\rep{1})  
\\ 
(-\tfrac 12,\tfrac 12,\tfrac12,-\tfrac 12,\tfrac 12, \tfrac 12,\pm\tfrac12^2)  &
&&&&&&&&& 
&  
\\ \hdashline
(0,-1,0,0,-1,0, 0^2)  & 
-\tfrac 12 & 1 & -\tfrac 12 & \tfrac 34 & -\tfrac 14 & \tfrac 34 & \tfrac 34 & 0 & -1 & 0
& (\crep{3},\rep{1}) 
\\ 
(\tfrac 12,-\tfrac 12,\tfrac12,\tfrac 12,-\tfrac 12, \tfrac 12,\pm\tfrac12^2)  &
&&&&&&&&& 
&   
\\ \hdashline 
(0,0,-1,0,0,-1, 0^2)  & 
-\tfrac 12 & -\tfrac 12 & 1 & \tfrac 34 & \tfrac 34 & -\tfrac 14 & \tfrac 34 & 0 & 0 & -1
& (\crep{3},\rep{1}) 
\\ 
(\tfrac 12,\tfrac 12,-\tfrac12,\tfrac 12,\tfrac 12, -\tfrac 12,\pm\tfrac12^2)  &
&&&&&&&&& 
& 
\\ \hline 
(0,0,0,0,1,1, 0^2)  & 
0 & -1 & -1 & -\tfrac 12 & \tfrac 12 & \tfrac 12 & \tfrac 12 & -1 & 0 & 0
& (\rep{3},\rep{1})
\\
(-\tfrac 12,-\tfrac 12,-\tfrac12,-\tfrac 12,\tfrac 12, \tfrac 12,\pm\tfrac12^2)  &
&&&&&&&&& 
&   
\\ \hdashline 
(0,0,0,1,0,1, 0^2)  & 
-1 & 0 & -1 & \tfrac 12 & -\tfrac 12 & \tfrac 12 & \tfrac 12 & 0 & -1 & 0
& (\rep{3},\rep{1}) 
\\ 
(-\tfrac 12,-\tfrac 12,-\tfrac12,\tfrac 12,-\tfrac 12, \tfrac 12,\pm\tfrac12^2)  &
&&&&&&&&& 
&  
\\ \hline 
}
& 
\arry{|c||ccc|cccc|ccc|c|}{
\hline 
\text{weight} & \mathsf{H}_1 & \mathsf{H}_2 & \mathsf{H}_3 & 
\mathsf{N}^1 & \mathsf{N}^2 &\mathsf{N}^3 &\mathsf{N}^{S} &
\gD\mathsf{N}^{1} & \gD\mathsf{N}^{2} & \gD\mathsf{N}^{3} & 
\text{repr.}
\\ \hline\hline 
(0,0,0,1,1,0, 0^2)  & 
-1 & -1 & 0 & \tfrac 12 & \tfrac 12 & -\tfrac 12 & \tfrac 12 & 0 & 0 & -1
& (\rep{3},\rep{1}) 
\\ 
(-\tfrac 12,-\tfrac 12,-\tfrac12,\tfrac 12,\tfrac 12, -\tfrac 12,\pm\tfrac12^2)  &
&&&&&&&&& 
&   
\\ \hline 
(-\tfrac 12,-\tfrac 12,-\tfrac12,\tfrac 12,\tfrac 12, \tfrac 12, \undr{-\tfrac 12, \tfrac 12})  & 
-1 & -1 & -1 & \tfrac 54 & \tfrac 54 & \tfrac 54 & \tfrac 54 & 0 & 0 & 0
& (\rep{1},\rep{2})  
\\ \hline 
(-\tfrac 12,-\tfrac 12,\tfrac12,-\tfrac 12,\tfrac 12, \tfrac 12, \undr{-\tfrac 12, \tfrac 12})  &
\tfrac 12 & -\tfrac 12 & -1 & -\tfrac 34 & \tfrac 14 & \tfrac 14 & \tfrac 14 & -1 & 0 & 0
& (\rep{1},\rep{2})  
\\ \hdashline 
(-\tfrac 12,\tfrac 12,-\tfrac12,-\tfrac 12,\tfrac 12, \tfrac 12, \undr{-\tfrac 12, \tfrac 12})  &
\tfrac 12 & -1 & -\tfrac 12 & -\tfrac 34 & \tfrac 14 & \tfrac 14 & \tfrac 14 & -1 & 0 & 0
& (\rep{1},\rep{2})  
\\ \hdashline 
(-\tfrac 12,-\tfrac 12,\tfrac12,\tfrac 12,-\tfrac 12, \tfrac 12, \undr{-\tfrac 12, \tfrac 12})  & 
-\tfrac 12 & \tfrac 12 & -1 & \tfrac 14 & -\tfrac 34 & \tfrac 14 & \tfrac 14 & 0 & -1 & 0
& (\rep{1},\rep{2})  
\\ \hdashline 
 (\tfrac 12,-\tfrac 12,-\tfrac12,\tfrac 12,-\tfrac 12, \tfrac 12, \undr{-\tfrac 12, \tfrac 12})   &
-1 & \tfrac 12 & -\tfrac 12 & \tfrac 14 & -\tfrac 34 & \tfrac 14 & \tfrac 14 & 0 & -1 & 0
& (\rep{1},\rep{2})  
\\ \hdashline 
 (-\tfrac 12,\tfrac 12,-\tfrac12,\tfrac 12,\tfrac 12,-\tfrac 12, \undr{-\tfrac 12, \tfrac 12})  & 
-\tfrac 12 & -1 & \tfrac 12 & \tfrac 14 & \tfrac 14 & -\tfrac 34 & \tfrac 14 & 0 & 0 & -1
& (\rep{1},\rep{2})  
\\ \hdashline 
(\tfrac 12,-\tfrac 12,-\tfrac12,\tfrac 12,\tfrac 12, -\tfrac 12, \undr{-\tfrac 12, \tfrac 12})  & 
-1 & -\tfrac 12 & \tfrac 12 & \tfrac 14 & \tfrac 14 & -\tfrac 34 & \tfrac 14 & 0 & 0 & -1
& (\rep{1},\rep{2})  
\\  \hline 
(-1,0,0,1,0,0, 0^2)  & 
-1 & -\tfrac 12 & -\tfrac 12 & \tfrac 14 & \tfrac 14 & \tfrac 14 & \tfrac 14 & 0 & 0 & 0
& (\rep{1},\rep{1})
\\ \hdashline 
(0,-1,0,0,1,0, 0^2)  & 
-\tfrac 12 & -1 & -\tfrac 12 & \tfrac 14 & \tfrac 14 & \tfrac 14 & \tfrac 14 & 0 & 0 & 0
& (\rep{1},\rep{1}) 
\\ \hdashline 
(0,0,-1,0,0,1, 0^2)  & 
-\tfrac 12 & -\tfrac 12 & -1 & \tfrac 14 & \tfrac 14 & \tfrac 14 & \tfrac 14 & 0 & 0 & 0
& (\rep{1},\rep{1})
\\ \hline 
(1,0,0,0,-1,0, 0^2)  & 
0 & \tfrac 32 & \tfrac 12 & 1 & 0 & 0 & 0 & 1 & 0 & 0
& (\rep{1},\rep{1})  
\\  \hdashline 
(1,0,0,0,0,-1, 0^2)  & 
0 & \tfrac 12 & \tfrac 32 & 1 & 0 & 0 & 0 & 1 & 0 & 0
& (\rep{1},\rep{1})  
\\  \hdashline 
(0,1,0,-1,0,0, 0^2)  & 
\tfrac 32 & 0 & \tfrac 12 & 0 & 1 & 0 & 0 & 0 & 1 & 0
& (\rep{1},\rep{1})  
\\  \hdashline 
(0,1,0,0,0,-1, 0^2)  & 
\tfrac 12 & 0 & \tfrac 32 & 0 & 1 & 0 & 0 & 0 & 1 & 0
& (\rep{1},\rep{1})  
\\  \hdashline 
(0,0,1,-1,0,0, 0^2)  & 
\tfrac 32 & \tfrac 12 & 0 & 0 & 0 & 1 & 0 & 0 & 0 & 1
& (\rep{1},\rep{1})  
\\  \hdashline 
(0,0,1,0,0,-1, 0^2)  & 
\tfrac 12 & \tfrac 32 & 0 & 0 & 0 & 1 & 0 & 0 & 0 & 1
& (\rep{1},\rep{1})  
\\ \hline 
(0,0,0,0,1,-1, 0^2)  & 
0 & -1 & -1 & -\tfrac 12 & \tfrac 12 & \tfrac 12 & \tfrac 12 & -1 & 0 & 0
& (\rep{1},\rep{1})  
\\  \hdashline 
(0,0,0,1,0,-1, 0^2)  & 
-1 & 0 & -1 & \tfrac 12 & -\tfrac 12 & \tfrac 12 & \tfrac 12 & 0 & -1 & 0
& (\rep{1},\rep{1})  
\\  \hdashline 
(0,0,0,1,-1,0, 0^2)  & 
-1 & -1 & 0 & \tfrac 12 & \tfrac 12 & -\tfrac 12 & \tfrac 12 & 0 & 0 & -1
& (\rep{1},\rep{1})  
\\ \hline
\multicolumn{12}{|c|}{ \phantom{xxx} }
\\[-1ex]  \hline 
}
\end{array}
\)}
\end{center} 
\caption{ \label{tb:SU32ChargedVectorStates} 
The line bundle charges $\mathsf{H}_i$, the triangulation multiplicities $\mathsf{N}^i$, $\mathsf{N}^S$ and the triangulation difference multiplicities $\gD\mathsf{N}^i$ are given for the $SU(3)\times SU(2)$ charged and singlet states obtained from the line bundle background defined by~\eqref{eq:SU32Bundles}.
}
\end{sidewaystable}

Since in this model all the gauge flux is located in a single $E_8$, the loop--corrected DUY equations in the form~\eqref{eq:SingleE8DUYs} can be used. Since for this model there are only three bundle vectors $\cV_1$, $\cV_2$ and $\cV_3$ which are clearly independent, the DUY equations reduce to three equations: 
\equ{
\arry{rcl}{ 
\sum\limits_{\gb\gg} e^{-2\gf}\, \text{Vol}(E_{1,\gb\gg}) &=& \dfrac{1}{16\gp} \big(64 + 2\,N_{E_1} - 2\,N_{E_2} - 2\,N_{E_3}\big)~, 
\\[2ex] 
\sum\limits_{\ga\gg} e^{-2\gf}\, \text{Vol}(E_{2,\ga\gg}) &=& \dfrac{1}{16\gp} \big(64 + 2\,N_{E_2} - 2\,N_{E_1} - 2\,N_{E_3}\big)~, 
\\[2ex] 
\sum\limits_{\gb\gg} e^{-2\gf}\, \text{Vol}(E_{3,\ga\gb}) &=& \dfrac{1}{16\gp} \big(64 + 2\,N_{E_3} - 2\,N_{E_1} - 2\,N_{E_2}\big)~. 
}
}
Since, the sum of volumes all need to be non--negative, the right--hand--sides of these equations all have to be positive. This leads to the conditions on the number of times the exceptional triangulations may be chosen: 
\equ{
N_{E_2} + N_{E_3} - N_{E_1} \leq 32~, 
\quad 
N_{E_1} + N_{E_3} - N_{E_2} \leq 32~, 
\quad 
N_{E_1} + N_{E_2} - N_{E_3} \leq 32~.  
}
Adding two of these three conditions shows that $N_{E_i} \leq 32$. In addition,  \eqref{eq:ConstraintTriangulationStepFunctions} implies that 
\equ{  
N_{E_1} + N_{E_2} + N_{E_3} \leq 64~. 
}
Thus apparently, one can only choose the $S$--triangulation at all 64 resolved singularities, but not one of the exceptional triangulations. However, one can choose to use exceptional triangulations at all resolved singularities, but not at all of them the same one. For example, the choice, $N_{E_1}=N_{E_2}=16$ and $N_{E_3}=32$, would be allowed by the DUY conditions.

\section{Jumping Spectra in a Blasczcyk--like GUT model}
\label{sc:BlasczcykGUT}

In ref.\cite{Blaszczyk:2010db} a semi--realistic MSSM model line bundle model on a resolution of $T^6/\Intr_2\times\Intr_2$ was constructed with gauge group $\mathrm{SU}(5) \times \mathrm{SU}'(3) \times \mathrm{SU}'(2)$. This model possessed an freely acting involution that reduced the gauge symmetry to the standard model gauge group. For this model the $E_1$--triangulation was chosen at all 64 resolved $\Cplx^3/\Intr_2\times \Intr_2$. In this section models similar to the Blasczcyk's GUT model are considered. The emphasis is not so much on finding a phenomenologically satisfactory model but rather on illustrating the effects of flop--transitions on the spectrum.

\subsection{Generalities of Blasczcyk--like GUT models}

Models like the Blaszczyk's GUT model are particular resolution of an orbifold theory with, in addition to two shifts $V_1$ and $V_2$ associated to the twists $v_1$ and $v_2$, up to five Wilson lines in all torus directions are switched on. The Wilson lines in the second, fourth and sixth torus directions are all taken equal: $W_2=W_4=W_6$ and independent of the two remaining Wilson lines $W_3$ and $W_5$. The resulting line bundle vectors are given by 
\equ{ \label{eq:DefBundleVectorsBlaszczyk}
\arry{rcl}{
\cV_{1,\gb\gg} = \cV_{1\gb_3\gg_5(\gb_4+\gg_6)} &=& 
V_1 + \gb_3\, W_3 + \gg_5\, W_5 + (\gb_4+\gg_6) W_2 + L_{1\gb_3\gg_5(\gb_4+\gg_6)} 
\\[1ex] 
\cV_{2,\ga\gg} = \cV_{2\_\gg_5(\ga_2+\gg_6)} &=& 
V_1 +  \gg_5\, W_5 + (\ga_2+\gg_6) W_2 + L_{2\gg_5(\ga_2+\gg_6)} 
\\[1ex] 
\cV_{3,\ga\gb} = \cV_{3\gb_3\_(\ga_2+\gb_4)} &=& 
V_3 + \gb_3\, W_3 +  (\ga_2+\gb_4) W_2 + L_{3\gg_5(\ga_2+\gb_4)} 
}
}
using the binary multi--index notation introduced in Subsection~\ref{sc:T6Z22orbifold}. Here 
$L_{1\gb_3\gg_5(\gb_4+\gg_6)} $, $L_{2\gg_5(\ga_2+\gg_6)} $ and $L_{3\gg_5(\ga_2+\gb_4)} $ are appropriately chosen $E_8\times E_8$ lattice vectors. The sum in between brackets is defined modulo two (since two times a Wilson line is a lattice vector which can be absorbed in one of the $L$'s). Thus, in total these kind of blowup models are defined by $8+4+4=16$ line bundle vectors and the 64 resolved fixed points are distinguished in 32 bunches of two fixed points as the index $\ga_1=0,1$ still parameterises a twofold degeneracy. In addition, there is a freely acting symmetry in such models: if one simultaneously adds 1 to the three indices $\ga_2, \gb_4, \gg_6$ modulo two: 
\equ{
(\ga_2, \gb_4, \gg_6) \mapsto (\ga_2+1,\gb_4+1,\gg_6+1)~,  
}
all bundle vectors are identical. 
This isometry was used in ref.\cite{Blaszczyk:2010db} introduce a freely acting Wilson line to break the $SU(5)$ GUT to the standard model. This step won't be considered here.

\begin{table}
\[
\arry{|c|c|}{
\hline 
\multicolumn{2}{|c|}{\text{Bundle vectors}} 
\\ \hline\hline 
\cV_{1000} = \cV_{1010} & 
(-\sfrac12, -\sfrac12, 1, 0, 0, 0, 0, 0)(0, 0, 0, 0, 0, 0, 0, 0)
\\ 
\cV_{1100}  = \cV_{1110} & 
(0, 1, 0, 0, 0, 0, 0, 0)( 0, 0, 0, 0, 0, 0, -\sfrac12, -\sfrac12)
\\
\cV_{1001} = \cV_{1011} & 
(\sfrac 14, \sfrac14, \sfrac34, -\sfrac14, -\sfrac14, -\sfrac14, -\sfrac14, -\sfrac14)(-\sfrac14, -\sfrac14, -\sfrac14, -\sfrac14, -\sfrac14, -\sfrac14, \sfrac14, \sfrac14)
 \\
\cV_{1101} = \cV_{1111} & 
(\sfrac14, -\sfrac34, \sfrac14, \sfrac14, \sfrac14, \sfrac14, \sfrac14, \sfrac14)( -\sfrac14, -\sfrac14, -\sfrac14, -\sfrac14, -\sfrac14, -\sfrac14, -\sfrac14, -\sfrac14)
\\ \hline 
\cV_{2\_00} = \cV_{2\_10} & 
(\sfrac14, -\sfrac14, \sfrac14, -\sfrac14, -\sfrac14, -\sfrac14, -\sfrac14, -\sfrac14)( 0, 0, 0, 0, 0, 0, 0, -1)
\\ 
\cV_{2\_01} = \cV_{2\_11} & 
(\sfrac12, 0, \sfrac12, 0, 0, 0, 0, 0)( \sfrac14, \sfrac14, \sfrac14, \sfrac14, -\sfrac34, \sfrac14, -\sfrac14, -\sfrac14)
\\ \hline  
\cV_{30\_0} & 
(-\sfrac14, \sfrac14, \sfrac14, -\sfrac14, -\sfrac14, -\sfrac14, -\sfrac14, -\sfrac14)( 0, 0, 0, 0, 0, 0, -1, 0)
\\
\cV_{31\_0} & 
(-\sfrac14, \sfrac14, \sfrac34, \sfrac14, \sfrac14, \sfrac14, \sfrac14, \sfrac14)( 0, 0, 0, 0, 0, 0, \sfrac12, -\sfrac12)
\\
\cV_{30\_1} &
(0, \sfrac12, \sfrac12, 0, 0, 0, 0, 0)( \sfrac14, \sfrac14, \sfrac14, \sfrac14, \sfrac14, -\sfrac34, -\sfrac14, -\sfrac14)
\\
\cV_{31\_1} &
(-\sfrac12, 0, \sfrac12, 0, 0, 0, 0, 0)( -\sfrac14, -\sfrac14, -\sfrac14, -\sfrac14, -\sfrac14, \sfrac34, -\sfrac14, -\sfrac14)
\\ \hline 
}
\]
\caption{A set of bundle vectors associated to two shifts and four Wilson lines that satisfy the flux quantisation conditions and the Bianchi identities in all triangulations.
\label{tb:BundleVectorsBlasczcyk} }
\end{table}

\subsection{Triangulation independent Blaszczyk--like GUT models}

The aim of this section is to engineer a modification of the Blaszczyk's GUT model such that  it fulfils the Bianchi identities in an arbitrary triangulation. As this turned out to be a very difficult, here only models are considered in which the Wilson lines $W_2=W_4=W_6$ and $W_3$ are switched on. Concretely the orbifold data of the model under consideration here is given by: 
\equ{
\arry{c}{
V_1= (\sm\sfrac 12, \sm\sfrac 12, 1, 0^5)(0^8)~, 
\qquad 
V_2 = (\sfrac 14, \sm\sfrac 14,\sfrac 14,\sm\sfrac 14^5)(0^6,0,\sm 1)~, 
\\[2ex]
W_3 = (0,0,\sfrac 12, \sfrac 12^5)(0^6,\sm\sfrac 12,\sm\sfrac 12)~, 
\qquad 
W_2=W_4=W_6 = (\sfrac 14, \sfrac 14, \sfrac 14, \sfrac 14^5)(\sm\sfrac 14^6,\sfrac 14,\sfrac 14)~. 
}
}
Using the freedom to add lattice vectors in~\eqref{eq:DefBundleVectorsBlaszczyk} it is possible to obtain a set of bundle vectors that satisfy the strong conditions~\eqref{eq:TriangulationFluxQuantisation}and~\eqref{eq:BundleVectorsTwistedStates}, 
which guarantee that the flux quantisation conditions and the Bianchi identities are satisfied in any triangulation. Such a set is given in Table~\ref{tb:BundleVectorsBlasczcyk}.

\begin{table}
\begin{center} 
\scalebox{1}{\(
\renewcommand{\arraystretch}{1.25} 
\arry{|cc|cc||cc|cc|}{
\hline 
\multicolumn{4}{|c||}{\text{Observable }E_8} & \multicolumn{4}{c|}{\text{Hidden }E_8}  
\\ \hline\hline 
\multicolumn{2}{|c|}{\text{$SU(5)$--adjoint}} & \multicolumn{2}{c||}{\text{Singlets}} & 
\multicolumn{2}{c|}{\text{$SU(4)$--adjoint}} & \rep{1}_4 & (0^4,0,1,0,1) 
\\
\rep{24} & (0^3,\undr{1,-1,0^3})  & \rep{1}_1 & (1,1,0,0^5) &
\rep{15} & (\undr{1,-1,0^2},0^4) &  \rep{1}_5 & (0^4,0,0,1,1) 
\\\cline{1-2} \cline{5-6}  
\multicolumn{2}{|c|}{\text{5--plets}} & \rep{1}_2 & (1,0,1,0^5) &
\multicolumn{2}{c|}{\text{4--plets}} &  \rep{1}_6 & (0^4,1,\sm1,0,0) 
\\ 
\rep{5}_1 & (0,1,0,\undr{1,0^4}) & \rep{1}_3 & (0,1,1,0^5) & 
\rep{4}_1 & (\undr{1,0^3},0,0,1,0) & \rep{1}_7 & (0^4,1,0,\sm1,0) 
\\  
\rep{5}_2 & (0,0,1,\undr{1,0^4}) & \rep{1}_4 & (1,\sm1,0,0^5) & 
\rep{4}_2 & (\undr{1,0^3},0,0,0,1)  &  \rep{1}_8 & (0^4,1,0,0,\sm1) 
\\ 
\rep{5}_3 & (0,\sm1,0,\undr{1,0^4}) & \rep{1}_5 & (1,0,\sm1,0^5) & 
\rep{4}_3 & (\undr{1,0^3},\sm1,0,0,0) &  \rep{1}_8 & (0^4,1,0,0,\sm1) 
\\ 
\rep{5}_4 & (0,0,\sm1,\undr{1,0^4}) & \rep{1}_6 & (0,1,\sm1,0^5) & 
\rep{4}_4 & (\undr{1,0^3},0,\sm1,0,0) & \rep{1}_{10} & (0^4,0,1,0,\sm1)  
\\ 
\rep{5}_5 & (\sm\sfrac 12, \sm\sfrac 12, \sfrac 12, \undr{\sfrac 12, \sm\sfrac 12^4}) & \rep{1}_7 & (\sm\sfrac 12, \sm\sfrac 12, \sfrac 12,  \sfrac 12^5) & 
\rep{4}_5 & (\undr{1,0^3},0,0,\sm1,0) & \rep{1}_{11} & (0^4,0,0,1,\sm1) 
\\ 
\rep{5}_6 & (\sm\sfrac 12, \sfrac 12, \sm\sfrac 12, \undr{\sfrac 12, \sm\sfrac 12^4}) & \rep{1}_8 & (\sm\sfrac 12, \sfrac 12, \sm\sfrac 12,  \sfrac 12^5) & 
\rep{4}_6 & (\undr{1,0^3},0,0,0,\sm1) & \rep{1}_{12} & (\sfrac 12^4,\sfrac 12,\sfrac 12, \sfrac 12,\sfrac 12) 
\\ 
\rep{5}_7 & (\sfrac 12, \sm\sfrac 12, \sm\sfrac 12, \undr{\sfrac 12, \sm\sfrac 12^4}) &  \rep{1}_9 & (\sfrac 12, \sm\sfrac 12, \sm\sfrac 12,  \sfrac 12^5) & 
\rep{4}_7 & (\undr{\sfrac 12,\sm\sfrac 12^3},\sm\sfrac 12,\sfrac 12,\sm\sfrac 12,\sm\sfrac 12) & \rep{1}_{13} & (\sfrac 12^4,\sm\sfrac 12,\sm\sfrac 12, \sfrac 12,\sfrac 12) 
\\ 
\rep{5}_8 & (\sfrac 12, \sfrac 12, \sfrac 12, \undr{\sfrac 12, \sm\sfrac 12^4}) & \rep{1}_{10} & (\sfrac 12, \sfrac 12, \sfrac 12,  \sfrac 12^5) & 
\rep{4}_8 & (\undr{\sfrac 12,\sm\sfrac 12^3},\sm\sfrac 12,\sfrac 12,\sfrac 12,\sfrac 12) &  \rep{1}_{14} & (\sfrac 12^4,\sm\sfrac 12,\sfrac 12, \sm\sfrac 12,\sfrac 12) 
\\ \cline{1-4} 
\multicolumn{2}{|c}{\text{10--plet}} & \rep{10} & (\sfrac 12, \sfrac 12, \sm\sfrac 12, \undr{\sfrac 12^2, \sm\sfrac 12^3}) &  \rep{4}_9 & (\undr{\sfrac 12,\sm\sfrac 12^3},\sfrac 12,\sm\sfrac 12,\sfrac 12,\sfrac 12) & \rep{1}_{15} & (\sfrac 12^4,\sm\sfrac 12,\sfrac 12, \sfrac 12,\sm\sfrac 12)  
\\  \cline{1-6} 
\multicolumn{4}{c||}{} &  \multicolumn{2}{c|}{\text{Singlets}} & \rep{1}_{16} & (\sfrac 12^4,\sfrac 12,\sm\sfrac 12, \sm\sfrac 12,\sfrac 12) 
\\ 
\multicolumn{4}{c||}{} & \rep{1}_1 & (0^4,1,0,1,0) & \rep{1}_{17} & (\sfrac 12^4,\sfrac 12,\sm\sfrac 12, \sfrac 12,\sm\sfrac 12) 
\\ 
\multicolumn{4}{c||}{} & \rep{1}_2 & (0^4,1,0,0,1) & \rep{1}_{18} & (\sfrac 12^4,\sfrac 12,\sfrac 12, \sm\sfrac 12,\sm\sfrac 12) 
\\ 
\multicolumn{4}{c||}{} & \rep{1}_3 & (0^4,0,1,1,0) &  \rep{1}_{19} & (\sfrac 12^4,\sm\sfrac 12,\sm\sfrac 12,\sm\sfrac 12,\sm\sfrac 12) 
\\ \cline{5-8} 
}
\)}
\end{center} 
\renewcommand{\arraystretch}{1} 
\caption{The identification between the roots and the states in the spectrum in both the observable and hidden sectors. States in the same non--Abelian representation but with different $U(1)$--charges are enumerated. 
\label{tb:States}}
\end{table}

\clearpage

\begin{table}[h!t]
\begin{center} 
\scalebox{0.7}{\(
\renewcommand{\arraystretch}{1.6} 
\arry{|c||c|c|c|c|}{
\hline 
\text{Resolved} & \text{Spectra in S--triangulation} & \multicolumn{3}{c|}{\text{Spectrum jumps due to flop--transitions}} 
\\
\text{fixed points} & 4 \times \mathsf{N}^{S} & \gD\mathsf{N}^{1}  & \gD\mathsf{N}^{2} & \quad\qquad \gD\mathsf{N}^{3} \quad\qquad\qquad 
\\ \hline\hline 
f_{\ga_10\,00\,\gg_50}\,,  & 
 \rep{5}_2 +\rep{5}_4 +  \crep{5}_5 +  \crep{10} + \crep{1}_1 +  3\,\rep{1}_7 +  \rep{1}_8 +  \rep{1}_9 & 
\crep{1}_7 & \crep{1}_5 & \rep{1}_6
\\ 
f_{\ga_11\,01\,\gg_51}  &
\cellcolor{gray!20} \crep{4}_1 + \crep{4}_2 + \rep{4}_5 + \rep{4}_6 +
\crep{1}_1 + \crep{1}_2 + \crep{1}_3 + \crep{1}_4 + 2\, \rep{1}_5 + \rep{1}_{7} + \rep{1}_{8} + &
\cellcolor{gray!20} \crep{1}_5 & \cellcolor{gray!20} \crep{1}_{11} & \cellcolor{gray!20} \rep{1}_{11}  
\\ 
& \cellcolor{gray!20}  \rep{1}_{9} + \rep{1}_{10} &\cellcolor{gray!20} & \cellcolor{gray!20}& \cellcolor{gray!20}
\\\hline 
f_{\ga_10\,10\,\gg_50}\,,~ & 
\crep{5}_1 + \rep{5}_2 + \crep{5}_3 + \rep{5}_7 + \rep{1}_1 + 2\, \crep{1}_3 + \rep{1}_4 + \crep{1}_5 + 3\, \rep{1}_6 + 3\, \rep{1}_7 +  &
\crep{1}_6 & \rep{5}_1+\crep{5}_7 + \rep{1}_3  + \crep{1}_4 +  & \crep{1}_7
\\ f_{\ga_11\,11\,\gg_51} &  3\, \crep{1}_8 + \crep{1}_{10} & & \rep{1}_8 + \rep{1}_{10} & 
\\
 &
\cellcolor{gray!20} \rep{4}_2+\crep{4}_6 + \rep{1}_2 + \rep{1}_4 + 2\, \rep{1}_5 + \crep{1}_8 + \crep{1}_{10} + 2\, \crep{1}_{11} &
\cellcolor{gray!20} \rep{1}_{11} & \cellcolor{gray!20}  & \cellcolor{gray!20} \crep{1}_5
\\\hline 
f_{\ga_11\,00\,\gg_50}\,, & 
\crep{5}_2 + \rep{5}_4 + \rep{5}_5 + \crep{10} 
+3\, \crep{1}_1 + \crep{1}_2 + \crep{1}_3 + \rep{1}_7 &
\rep{1}_1 & \crep{1}_5 & \crep{1}_6
\\
f_{\ga_10\,01\,\gg_51} &
\cellcolor{gray!20} \rep{4}_3 + \rep{4}_4 + \crep{4}_8 + \crep{4}_9 + 
\crep{1}_1 + \crep{1}_2 + \crep{1}_3 + \crep{1}_4 + 
\rep{1}_{14} +   \rep{1}_{15} + \rep{1}_{16} +  &
\cellcolor{gray!20} \rep{1}_{19} & \cellcolor{gray!20} \rep{1}_6 & \cellcolor{gray!20} \crep{1}_6 
\\ 
& \cellcolor{gray!20}\rep{1}_{17} + 2\, \crep{1}_{19} & \cellcolor{gray!20} & \cellcolor{gray!20} & \cellcolor{gray!20}
\\\hline 
f_{\ga_11\,10\,\gg_50}\,, & 
\rep{5}_1 + \crep{5}_3 + 
\crep{1}_1 + \rep{1}_2 + \crep{1}_3 + \rep{1}_4 + \crep{1}_5 + 3\, \rep{1}_6  &
\crep{1}_6 & \crep{1}_4 & \rep{1}_1
\\
f_{\ga_10\,11\,\gg_51} &
\cellcolor{gray!20} \rep{4}_3 + \crep{4}_4 + \crep{4}_7 + \rep{4}_9 + 
\rep{1}_1 + \rep{1}_2 + \rep{1}_5 + 2\, \rep{1}_6 + \crep{1}_9 +  \crep{1}_{10} + \rep{1}_{13} +   &
\cellcolor{gray!20} \crep{1}_6 & \cellcolor{gray!20} \crep{1}_{13} & \cellcolor{gray!20} \rep{1}_{19} 
\\ 
& \cellcolor{gray!20} \rep{1}_{14} + \rep{1}_{15}  + \crep{1}_{16} + \crep{1}_{17} + \crep{1}_{19} & \cellcolor{gray!20} & \cellcolor{gray!20} & \cellcolor{gray!20}
\\\hline 
f_{\ga_10\,01\,\gg_50}\,, & 
\rep{5}_4 + \crep{5}_8 + \crep{1}_2 + 2\,\crep{1}_3 + 2\,\rep{1}_8 + \rep{1}_9 &
 & \rep{1}_3 & \crep{1}_8 
\\ 
f_{\ga_11\,00\,\gg_51}  & 
\cellcolor{gray!20} \crep{4}_2 + \rep{4}_4 + \crep{4}_6 + \rep{4}_8 + 
\crep{1}_2 + \crep{1}_3 + 2\, \rep{1}_4 +\rep{1}_5 + \rep{1}_6 + 2\, \crep{1}_8 + 3\, \rep{1}_{10} + & 
\cellcolor{gray!20} \rep{4}_6+ \crep{4}_8 + 
\crep{1}_4 + \crep{1}_5 +  & \cellcolor{gray!20} \crep{1}_{10} & \cellcolor{gray!20} \crep{1}_{16} 
\\ 
& \cellcolor{gray!20}   \rep{1}_{11} + 3\, \rep{1}_{16} + \crep{1}_{17} + 3\, \crep{1}_{18} + \crep{1}_{19} & \cellcolor{gray!20} \rep{1}_8 + \rep{1}_{17} + \rep{1}_{18} + \rep{1}_{19} & \cellcolor{gray!20} & \cellcolor{gray!20} 
\\\hline 
f_{\ga_10\,11\,\gg_50}\,, & 
\rep{5}_3 + \crep{5}_6 + \crep{1}_4 + \crep{1}_5 + \rep{1}_6 + \crep{1}_7 + \crep{1}_8 + 3\, \rep{1}_9 &
\crep{1}_9 & \rep{1}_7 & \rep{1}_4 
\\ 
f_{\ga_11\,10\,\gg_51} &
\cellcolor{gray!20} \rep{4}_2 + \crep{4}_4 + \rep{4}_6 + \crep{4}_7 + 
\rep{1}_2 + \rep{1}_4 + \rep{1}_5 + \crep{1}_6 + \rep{1}_8 + \rep{1}_9 + 2\, \crep{1}_{10} + & 
\cellcolor{gray!20} \rep{1}_{10} & \cellcolor{gray!20} \crep{1}_{17} & \cellcolor{gray!20} \crep{1}_4 
\\ 
& \cellcolor{gray!20}  \rep{1}_{11} + \rep{1}_{12} + \rep{1}_{13} + \rep{1}_{16} + \rep{1}_{17}  & \cellcolor{gray!20}  & \cellcolor{gray!20}  & \cellcolor{gray!20} 
\\\hline 
f_{\ga_10\,00\,\gg_51}\,, & 
\rep{5}_4 + \crep{5}_8 + 2\, \crep{1}_2 + \crep{1}_3 + \rep{1}_8 + 2\,\rep{1}_9  & 
 & \crep{1}_9 & \rep{1}_2 
\\ 
f_{\ga_11\,01\,\gg_50}  & 
\cellcolor{gray!20} \crep{4}_1 + \rep{4}_3 + \crep{4}_5 + \rep{4}_9 + 
2\, \rep{1}_1 + \crep{1}_2 + \crep{1}_3 + \rep{1}_5 + \crep{1}_6 + 3\, \rep{1}_7 + \crep{1}_9 +  & 
\cellcolor{gray!20} \rep{4}_5 + \crep{4}_9 + \crep{1}_1 + \crep{1}_5 + 
& \cellcolor{gray!20} \crep{1}_{15} & \cellcolor{gray!20} \crep{1}_7 
\\ 
& \cellcolor{gray!20}  \crep{1}_{11} + \crep{1}_{14} + 3\, \rep{1}_{15} + 3\, \crep{1}_{18} + \crep{1}_{19} & \cellcolor{gray!20} \rep{1}_9 + \rep{1}_{14} + \rep{1}_{18} + \rep{1}_{19}  & \cellcolor{gray!20}  & \cellcolor{gray!20}  
\\\hline 
f_{\ga_10\,10\,\gg_51}\,, & 
\crep{5}_2 + \rep{5}_3 + \rep{5}_6 + \crep{5}_7 + 
\crep{1}_2 + \crep{1}_3 + \crep{1}_4 + \crep{1}_5 + \rep{1}_6  + 2\, \crep{1}_7 + \rep{1}_9 + \crep{1}_{10}  &
\rep{1}_3 & \rep{1}_7 & \rep{1}_4 
\\
 f_{\ga_11\,11\,\gg_50} & 
\cellcolor{gray!20} \rep{4}_3 + \crep{4}_9 + 
\rep{1}_1 + \rep{1}_2 + \crep{1}_6 + \rep{1}_{11} + \crep{1}_{12} + 3\, \rep{1}_{14} + \crep{1}_{15} + \rep{1}_{19}  & 
\cellcolor{gray!20} \rep{1}_{15} & \cellcolor{gray!20} \crep{1}_{14} & \cellcolor{gray!20} \crep{1}_1  
\\\hline 
}
\)}
\end{center} 
\renewcommand{\arraystretch}{1} 
\caption{
Each big row corresponds to two sets of four resolved $\Cplx^3/\Intr_2\times\Intr_2$ fixed points labelled by $\ga_1, \gg_5=0,1$ (because their local bundle vectors are identical and thus so are their local spectra). The lines with the white background give the observable spectra resulting from the first $E_8$ and the lines with grey background the hidden spectrum from the second $E_8$. The charge states are labeled in Table~\ref{tb:States}. (Since all singlet are charged it make sense to talk about a singlet state or its conjugate.) The second column gives  the contributions at the four local resolved singularities using the S--triangulation combined. The columns $\gD\mathsf{N}^{1}$, $\gD\mathsf{N}^{2}$ and $\gD\mathsf{N}^{3}$ indicate the jumps in the spectra for a single resolved fixed point out of these sets of four singularities. 
\label{tb:JumpingSpectraII}}
\end{table}

The resulting spectra are given in Table~\ref{tb:JumpingSpectraII}. The states used in that table are defined in Table~\ref{tb:States} from the roots of both $E_8$--factors. Notice, that not all $E_8$--roots (up to conjugation) appear here; only the states, that have a non--vanishing multiplicity in the models defined here, are listed. The subscripts are used to distinguish states that have the same non--Abelian representation but different $U(1)$ charges. The second column gives the spectra from the local resolved singularities when the $S$--triangulation is used at all 64 of them. Since the labels $\ga_1, \gg_5=0,1$ are arbitrary, there will be a fourfold degeneracy in the spectrum, this is already taking into account in the table by multiplying the spectra in the $S$--triangulation by 4. The additional two--fold degeneracy due to the freely action symmetry is made apparent by giving two sets of four resolved singularities. It is not difficult to see that the full spectrum using the $S$--triangulation is free of non--Abelian anomalies. 

The final three columns of Table~\ref{tb:JumpingSpectraII} displays the jumps in the spectra when at a given singularity the $S$--triangulation is flopped to the triangulation $E_1$, $E_2$ or $E_3$. These are the jumps at a single resolved fixed point. It can be seen that in accordance with our general findings this jumps are always integral. Most jumps that occur in the spectra involve singlets only. At the resolved fixed points $f_{\ga_10\,10\,\ga_50}$ and $f_{\ga_11\,11\,\gg_50}$ a $\rep{5}$ and $\crep{5}$ pair appears during a flop from the $S$ to the $E_2$--triangulation. Similarly, a $\rep{4}$ and $\crep{4}$ pair appears at resolved fixed points $f_{\ga_10\,01\,\gg_50}$ and $f_{\ga_10\,00\,\gg_51}$. Thus, at most only non--Abelian vector--like pairs can arise during a flop transition.

\section{Conclusion}
\label{sc:Conclusion}

\subsection*{Summary}

This paper has been devoted to a specific problem which occurs in resolutions of certain toroidal orbifolds, namely that the resolutions of the local singularities is not unique at the topological level and therefore leads to an explosion of topologically distinct smooth geometries all associated to one and the same orbifold. As a concrete working example the focus was on the resolutions of a $T^6/\Intr_2\times \Intr_2$ orbifold which contains 64 $\Cplx^3/\Intr_2\times \Intr_2$ singularities, each of which admits four distinct resolutions encoded by different triangulations of their toric diagram. 

The key idea to overcome this complication is to use a parameterisation to keep track of the triangulations chosen at all resolved fixed points simultaneously. It turned out not to be very cumbersome to express the fundamental (self--)intersection numbers of the divisors of the resolution in terms of this data. Once the (self--)intersection numbers were determined, many derived objects can be computed without much more difficulty as determining them within a specific triangulation. In particular, we checked our procedure by computing the integrated third Chern class directly and confirmed that it equals 96 independently of any triangulation choice. We obtained expressions for the volumes of curves, divisors and the manifold as a whole for any possible choice of the triangulation of the 64 $\Intr_2\times\Intr_2$ singularities. In addition, we worked out some of the fundamental consistency conditions of line bundle models on the resolutions of the $T^6/\Intr_2\times \Intr_2$ like the flux quantisation conditions and the integrated Bianchi identities (which for simplicity were only considered without five branes). Even a tool which is often used to compute the chiral part of the spectrum, the multiplicity operator, could be determined once and for all for any choice of triangulation. 

Having written down the fundamental consistency conditions for any possible choice of triangulation, allowed for posing the question what conditions have to be enforced to ensure that they are satisfied for all possible triangulations simultaneously. It turned out that if the flux quantisation conditions are satisfied for a given specific choice of triangulation, they are, in fact, fulfilled for any configuration of triangulations: the flux quantisation conditions turned out to be triangulation independent. The superimposed integrated Bianchi identities reduced to much simpler requirements than those within any particular choice of triangulation. Moreover, they are quite reminiscent of some of the properties of shifted momenta of the blowup modes that induce the resolution from the orbifold perspective. 

These ideas and results were illustrated by a number of examples in the remainder of the paper. For simplicity, first line bundle models were considered, where the 48 line bundle vectors were chosen to be determined by three defining vectors. By computing spectra in all triangulations explicitly, it was confirmed that the full chiral spectra are always integral. We take this as a very strong crosscheck of the procedure outlined in this paper to parameterise all possible triangulations of the resolved singularities of the $T^6/\Intr_2\times \Intr_2$ orbifold. This was also checked explicitly in a variant of the Blasczczyk's GUT model with four Wilson lines of which three were set equal. The full spectrum computed in the $S$--triangulation everywhere is integral and free of non--Abelian anomalies. But also all the local difference multiplicities measuring the jumps in the local spectra at specific resolved singularities are always integral and free of non--Abelian anomalies (as the jumping spectra were all vector--like in this particular example).

\subsection*{Outlook}

This paper focussed on one particular $T^6/\Intr_2\times \Intr_2$ orbifold, it is to be expected that this procedure can also be applied to the other $T^6/\Intr_2\times \Intr_2$ orbifolds. In fact, applications do not stop there, for any orbifold for which the resolution of some of the local singularities is not unique, it may be applied. Table~\ref{tb:TriangulationDependenceOrbifolds} gives an overview of some toroidal orbifolds for which the triangulations of their local singularities are not unique and a naive estimate of the number of resolved geometries which therefore can be associated to that orbifold. (The numbers quoted in this table are upper limits: these orbifolds can be defined on different lattices on which the number of fixed points may be lower than the numbers indicated here.) Moreover, triangulation ambiguities do not only show up in toroidal orbifolds resolutions, also in other Calabi--Yau constructions they might be present. For example, some Calabi--Yaus in the Kreuzer--Skarke list obtained as hypersurfaces in toric varieties are not unique due to different triangulation choices~\cite{Kreuzer:2000xy,Altman:2014bfa}. One may therefore speculate whether similar methods may also be applied there.

\begin{table}
\begin{center} 
\begin{tabular}{|c||c|c|c|}
\hline 
Toroidal & Number of & Triangulations & Naive number \\[-1ex] 
orbifold & fixed points & per fixed point & of resolutions 
\\ \hline\hline 
$T^6/\Intr_\text{6--II}$ & $12$ & $5$ & $5^{12} \sim 10^8$ \\ 
$T^6/\Intr_\text{2}\times\Intr_\text{2}$ & $64$ & $4$ & $4^{64} \sim 10^{38}$ \\
$T^6/\Intr_\text{2}\times\Intr_\text{4}$ & $24$ & $16$ & $16^{24} \sim 10^{28}$ \\
$T^6/\Intr_\text{3}\times\Intr_\text{3}$ & $27$ & $79$ & $79^{27} \sim 10^{51}$ 
\\ \hline 
\end{tabular}
\end{center} 
\caption{\label{tb:TriangulationDependenceOrbifolds} 
Triangulation dependence and the naive number of resulting resolutions of toroidal orbifolds as can be inferred from the data in ref.~\cite{Lust:2006zh}. }
\end{table}

Another direction of research where we could imagine that the results of this paper might be beneficial are investigations of the spinor--vector duality on smooth geometries. The spinor--vector duality is a symmetry akin to mirror symmetry in the space of $(2,0)$ heterotic--string compactifications~\cite{Faraggi2007b,Faraggi:2007ms,CatelinJullien:2008pc,Angelantonj:2010zj,Faraggi:2011aw,Faraggi:2021fdr, Faraggi:2021yck}. It arises due to the exchange of Wilson line moduli rather than moduli of the internal compactified space and operates separately on each of the twisted sectors in $\Intr_2\times \Intr_2$ orbifolds~\cite{Faraggi:2007ms,Faraggi:2011aw}
and hence can be studied in vacua with a single $\Intr_2$ twist of the internal space and an additional freely acting $\Intr_2$ that operates as a Wilson line. Similar to mirror symmetry~\cite{Candelas:1990rm, Greene:1990ud}, which can be realised as an exchange of discrete torsion in a $\Intr_2\times \Intr_2$ toroidal orbifold~\cite{Vafa:1994}, the spinor--vector duality can be realised in terms of certain generalised discrete torsions~\cite{Faraggi2007b,Angelantonj:2010zj,Faraggi:2011aw}. Moreover, like the imprint of mirror symmetry on Calabi--Yau manifolds, the spinor--vector duality imprints can be explored in the effective field theory limit of smooth compactifications as was investigated in refs.~\cite{Faraggi:2021yck} and \cite{Faraggi:2021fdr} in six and five dimensions, respectively. To take these studies further to the resolutions of $\Intr_2\times \Intr_2$ orbifolds the present work is likely to be instrumental as it allows to study the required resolutions in general and not be hampered by focusing on a particular triangulation from the very beginning.

%
%
%
%
%
%
%

\appendix 
\def\theequation{\thesection.\arabic{equation}} 

\section{Some Chern Class Identities}
\label{sc:MathRelations}
\setcounter{equation}{0}

The total Chern class $c$ of a matrix $A$ is given by 
\equ{
c = \det (\Id + A)~.  
}
If $x_i$ denote the eigenvalues of $A$, then the total Chern class can be written as
\equ{
c = \prod (1 +x_i)~. 
}
Expanding the total Chern class $c$ in different powers of $x_i$ can be used to define the first, second and third Chern classes, $c_1$, $c_2$ and $c_3$, respectively. These Chern classes may be expressed as
\equ{
c_1 = \sum_i x_i~, 
\quad 
c_2 = \sum_{i<j} x_i x_j~, 
\quad 
c_3 = \sum_{i<j<k} x_i x_j x_k~. 
}
The second Chern class can be rewritten as: 
\equ{ \label{eq:c2intermediate} 
c_2 = \dfrac 12 \sum_{i\neq j} x_i x_j 
= \dfrac 12 \sum_{i,j} x_i x_j - \dfrac 12 \sum_i x_i^2 = \dfrac 12\, c_1^2 - \dfrac 12 \sum_i x_i^2~. 
}
In a similar spirit the third Chern class can also be rewritten. When inserting
\equ{
\sum_{i\neq j} x_i x_j^2 = \sum_{i,j} x_i x_j^2 - \sum_i x_i^3
}
in
\equ{ 
\Big( \sum_i x_i^3 \Big)^3 = \sum_i x_i^3 + 3 \sum_{i\neq j} x_i x_j^2 + \sum_{i\neq j \neq k} x_i x_j x_k~, 
}
one finds the identity 
\equ{ \label{eq:c3intermediate} 
c_1^3 = -2 \sum_i x_i^3 + c_1 \sum_j x_j^2 + 3!\, c_3~. 
}
If the first Chern class vanishes, {i.e.}\ $c_1=0$, the relations~\eqref{eq:c2intermediate} and~\eqref{eq:c3intermediate} imply that 
\equ{ \label{eq:Expansion23Chern}
c_2 = - \dfrac 12 \sum_i x_i^2~, 
\qquad 
c_3 = \dfrac 13 \sum_i x_i^3~. 
}

\bibliographystyle{paper}
{\small
\bibliography{paper}
}
\end{document}